\documentclass[manuscript]{acmart}
\usepackage[hide]{chato-notes}
\usepackage{url}
\usepackage{multirow}
\usepackage{graphicx}
\usepackage[table,xcdraw]{xcolor}
\usepackage{tabularray} 

\AtBeginDocument{%
  }

\setcopyright{none}             
\begin{document}

\title{Fueling Volunteer Growth: the case of Wikipedia Administrators}

\author{Eli Asikin-Garmager}
\affiliation{%
  \institution{Wikimedia Foundation}
  \city{San Francisco}
  \state{California}
  \country{USA}}
  \email{easikingarmager@wikimedia.org}

\author{Yu-Ming Liou}
\affiliation{%
  \institution{Wikimedia Foundation}
  \city{San Francisco}
  \state{California}
  \country{USA}}
\email{yliou@wikimedia.org}

\author{Caroline Myrick}
\affiliation{%
  \institution{Wikimedia Foundation}
  \city{San Francisco}
  \state{California}
  \country{USA}}
\email{cmyrick@wikimedia.org}

\author{Claudia Lo}
\affiliation{%
  \institution{Wikimedia Foundation}
  \city{San Francisco}
  \state{California}
  \country{USA}}
\email{clo@wikimedia.org}

\author{Diego Saez-Trumper}
\affiliation{%
  \institution{Wikimedia Foundation}
  \city{San Francisco}
  \state{California}
  \country{USA}}
  \email{dsaeztrumper@wikimedia.org}

\author{Leila Zia}
\affiliation{%
  \institution{Wikimedia Foundation}
  \city{San Francisco}
  \state{California}
  \country{USA}}
  \email{leila@wikimedia.org}

\authorsaddresses{} 

\begin{abstract}
Wikipedia administrators are vital to the platform's success, performing over a million administrative actions annually. This multi-method study
systematically analyzes adminship across 284 Wikipedia languages since 2018, revealing a critical two-sided trend: while over half of all Wikipedias show a net increase in administrators, almost two-thirds of highly active Wikipedias face decline. Our analysis, drawing from large-scale adminship log analysis, over 3000 surveys, and 12 interviews, reveals this decline is primarily driven by insufficient recruitment, not unusual attrition. We identify key barriers for potential administrators, including limited awareness, ambiguous requirements, a demanding selection process, and low initial interest. Recognizing that current administrators remain highly motivated and engaged, we propose actionable recommendations to strengthen recruitment pipelines and fuel Wikipedia administrator growth, crucial for Wikipedia's long-term sustainability.
\end{abstract}





\maketitle

\section{Introduction}
\label{sec:intro}


Every second, Wikipedia is edited an average of 5 times, and viewed approximately 6000 times\footnote{Numbers are computed based on 12-month total pageviews and total edits to Wikipedia as of 2025-09-11 \url{https://stats.wikimedia.org/\#/all-wikipedia-projects}}. This immense scale, fueled by hundreds of thousands of volunteer editors across over 300 languages, relies not only on volunteer editors who write encyclopedic knowledge but also on a critical group of trusted Wikipedia editors with extended rights: \textit{Wikipedia Administrators}. Administrators perform a range of activities on Wikipedia including but not limited to deleting and restoring Wikipedia articles, blocking disruptive users, and restricting editing access through page protection mechanisms. Through their work they ensure timely, verifiable encyclopedic knowledge. While readers, editors, and content have been extensively studied \cite{ren2023did,okoli2014wikipedia,moas2023automatic}, the dynamics of these vital extended-rights users remain largely unexplored. This study focuses on administrators due to their indispensable role in maintaining content integrity and community health, and their position as a foundational tier for other extended-rights roles such as Wikipedia stewards (users with complete access to the wiki interface on all Wikimedia projects).

\noindent\textbf{Background and objective}
Despite the critical role of administrators, existing research on Wikipedia has primarily focused on editor and reader motivations \cite{singer2017we,lemmerich2019world,antin2010readers,kuznetsov2006motivations,xu2015empirical,yang2010motivations}, with limited understanding of administrators beyond specific studies on English Wikipedia's Request for Adminship (RFA) process \cite{danescu_echoes_2012,danescu_politeness_2013}. Critically, these prior works often overlook cross-language administrator trends, the underlying drivers of adminship changes, and the specific barriers faced by potential candidates. Concerns about administrator numbers have also been raised intermittently within the Wikipedia community itself, yet a comprehensive, multi-language investigation has been lacking. This study aims to address these significant gaps by offering 
a large-scale, multi-method analysis of Wikipedia adminship and administrators, their trends, challenges, and motivations across a diverse set of language editions.



\noindent\textbf{Contributions and findings.} 
This study presents the first comprehensive, multi-language investigation into Wikipedia administrator dynamics, leveraging log data from 284 Wikipedias, over 3000 administrator surveys, and 12 semi-structured interviews. Our primary contributions are as follows:
\begin{itemize}
    \item We reveal a critical two-sided trend in administrator numbers: while over half of all Wikipedias show growth or stability since 2018, roughly two-thirds of highly active Wikipedias face decline (Section~\ref{sec:results-trends}).
    \item We demonstrate that this decline is primarily driven by insufficient recruitment, rather than unusual attrition, with administrator departure rates remaining largely stable (Section~\ref{sec:results-recruitment}).
    \item We identify key barriers to recruitment, including limited awareness of the role, ambiguous candidacy requirements, a demanding and stressful selection process (RFA), and low initial interest among potential administrators. We also highlight the crucial opportunity of current administrator motivation and engagement, and the pervasive challenge of conflict (Section~\ref{sec:results-challenges}).
    \item We provide actionable recommendations derived from our findings and community engagement to strengthen administrator recruitment pipelines and foster sustainable volunteer moderation, relevant for both Wikipedia communities and the broader HCI community (Section~\ref{sec:discussion}).
\end{itemize}

\section{Background}
\label{sec:background}
In this section we provide background knowledge for this study, and explain key concepts used in this paper. Technical definitions will be covered in Section \ref{sec:methodology}.

\subsection{What Wikipedia administrators do and why it matters}
Administrators on Wikipedia conduct social, governance, and technical work essential to the healthy operation of their Wikipedia language editions. They are granted extended user rights, enabling them to delete pages, block disruptive users, and restrict editing access via page protection\footnote{\url{https://en.wikipedia.org/wiki/Wikipedia:Protection_policy}}. These rights are primarily used to combat vandalism,\footnote{\url{https://en.wikipedia.org/wiki/Wikipedia:Vandalism}} spam, and other issues that threaten content quality and accuracy, thereby maintaining editorial integrity~\cite{mathieu_oneil_wikipedia_2011}.

Beyond direct moderation, administrators are deeply involved in the project's infrastructure and community. They develop and operate third-party tools, such as bots for anti-vandalism efforts~\cite{r_stuart_geiger_lives_2011}. Socially, they exhibit various leadership behaviors~\cite{zhu_identifying_2011}, form crucial knowledge-sharing networks among themselves~\cite{karczewska_knowledge_2024}, and, as active and involved users, model normative behavior for other editors~\cite{panciera_wikipedians_2009}. Furthermore, administrators play a significant role in the creation and discussion of Wikipedia policies~\cite{butler_dont_2008}.

Membership in the administrator group also qualifies users for other, more-restrictive user roles, in practice leveraging the social capital and trust that adminship confers. These specialized roles are essential for critical, higher-tier work across diverse Wikipedia language editions. Examples include: \textit{CheckUsers},\footnote{\url{https://meta.wikimedia.org/wiki/CheckUser_policy}} who access IP information for anti-vandalism efforts; \textit{Oversighters},\footnote{\url{https://meta.wikimedia.org/wiki/Oversight_policy}} capable of hiding sensitive edits and redacting private or libelous information from histories; and \textit{Stewards},\footnote{\url{https://meta.wikimedia.org/wiki/Stewards}} responsible for user rights management across all Wikipedias.

\subsection{Gaining and losing administrator rights}
The process of becoming a Wikipedia administrator begins with a nomination of a user, and the presentation of their qualifications to the broader community. This process is generally known as a Request for Adminship (RFA)\footnote{The list of documented RFA processes across many Wikipedia languages can be accessed via \url{https://www.wikidata.org/wiki/Q4048254}}, and usually takes place on a public page where community members can ask questions and express support or opposition for a candidate. The duration and specific rules of RFAs vary widely across different language Wikipedias. Outcomes are decided by consensus or voting, with a typical success threshold of around two-thirds support. Some Wikipedia editions grant admin rights indefinitely, while others have term limits requiring re-election.

\subsection{Removal of administrative rights}
Administrators can have their rights removed voluntarily by stepping down or involuntarily through a ``de-sysopping''\footnote{\url{https://simple.wiktionary.org/wiki/Wiktionary:Desysop_policy}} process initiated by the community. The procedures for involuntary removal vary widely: some Wikipedia editions hold formal votes requiring evidence of wrongdoing and community support, while others rely on higher-level committees of other community members to deliberate on de-sysopping requests. Many projects also have inactivity rules, where administrators who do not maintain a minimum level of activity can lose their status. In cases of removal, former admins may be allowed to reapply for adminship later, depending on community policies. 

\section{Related Work}
\label{sec:related-work}
While substantial research has explored the social dynamics of Wikipedia editors, investigations specifically focused on its administrators remain limited. This section reviews relevant prior work and highlights the specific gaps in the literature that our study addresses.

\subsection{Trends in online community administration}
Efforts to describe the work performed by volunteer community moderators (analogous to Wikipedia administrators) exist for platforms such as Reddit \cite{matias_going_2016,gilbert_cesspool_2020,trujillo2022make,habib2022exploring} and Twitch \cite{taylor_watch_2018,seering_who_2023,thach_visible_2024}. For instance, studies have developed models for categorizing moderator engagement through interviews with 56 volunteer moderators from Twitch, Reddit, and Facebook \cite{seering_moderator_2019}, or described visible and invisible moderation work on Reddit using private datasets of 900 moderators across 126 subreddits \cite{li_all_2022}. Unfortunately, the closed nature of most of these platforms' user activity data limits the public and academic reach of such studies.

Despite previous inquiry into work performed by volunteer community moderators on other platforms, academically rigorous studies on Wikipedia administrators are notably sparse. While early work in 2008 examined Wikipedia admin dynamics \cite{burke_taking_2008, burke_mopping_2008}, subsequent discussions of administration trends have largely been confined to on-platform publications~\cite{signpost_recent_2015, signpost_thirteen_2023} and reports by news media organizations~\cite{meyer_no_admins_2012,hansi_lo_wang_as_2012,ian_steadman_wikipedia_2012}. These informal analyses frequently advance a narrative of declining administrator figures, particularly with press attention peaking in 2012.

Crucially, despite these long-standing community concerns and widespread media attention, there remains a significant gap in the academic literature: to the best of our knowledge, no existing cross-language studies have systematically analyzed Wikipedia administrator logs and trends. Our study addresses this by gathering and analyzing administrator logs from 284 Wikipedia language editions since 2018. This comprehensive approach allows us to answer the critical question of whether Wikipedia is truly losing administrators and to uncover a range of findings about admin activity and dynamics across diverse communities.

\subsection{Barriers to potential administrators}
One prominent line of research investigates the administrator pipeline by studying English Wikipedia's Request for Adminship (RFA) dynamics and successful candidates, often through building predictive models for RFA success. These models typically examine factors such as editors' activity patterns (e.g., edit volume, diversity, talk-page participation)~\cite{jankowski-lorek_modeling_2013,picot-clemente_social_2015,burke_taking_2008} and the relationship between voter and candidate characteristics~\cite{leskovec_governance_2010}. While these studies offer insights into RFA as a model for group decision-making, they predominantly focus on English Wikipedia and lack analyses of administrator-specific activity beyond the RFA process itself.

A separate body of research examines Wikipedia's administrators through the lens of governance models~\cite{mathieu_oneil_wikipedia_2011,arazy_functional_2015}. These studies describe power dynamics within Wikipedia, at times highlighting a hierarchy despite the project's anti-hierarchical ethos~\cite{arazy_functional_2015}, or as a case study of an open and democratic critique of expert-based authoritative rule~\cite{mathieu_oneil_wikipedia_2011}. Related scholarship further investigates RFAs as a site of contested power relationships, applying politeness theory to compare candidate behaviors pre- and post-election, and between successful and unsuccessful candidates~\cite{danescu_echoes_2012,danescu_politeness_2013}. While these studies adjust for the confounding factor of \textit{interest in adminship}, our research explicitly aims to understand the determinants of such interest. This necessitates examining a broader population of ``potential administrators'': individuals who meet RFA candidacy criteria but whose interest in adminship is not yet known.

In summary, prior work has not extensively explored the identities, experiences, or attitudes of potential administrators, particularly concerning their knowledge of and interest in adminship. Our study addresses this by making a novel contribution: we survey over 2,000 potential admins (editors meeting formal RFA requirements on their respective language editions) to systematically understand their awareness, attitudes, and interest in Wikipedia adminship. This approach provides a deeper understanding of the RFA process's complex social dynamics, moving beyond analyses of linguistic changes or structures of power and governance to probe the foundational issue of recruitment.

\subsection{Motivation in volunteer moderation}
Volunteer moderators in online platforms, including Wikipedia administrators, are critical for maintaining community norms, resolving conflicts, and safeguarding content quality~\cite{grimmelmann2015virtues}. However, the demanding nature of this work often leads to emotional exhaustion and burnout. Research highlights the invisible emotional labor moderators perform—de-escalating tensions, absorbing abuse, and maintaining civility—which is essential yet often under-recognized and unsupported \cite{matias_civic_2019}. This aligns with findings that volunteer content moderators on platforms such as Facebook and Reddit frequently quit due to interpersonal conflict, burnout, and unsupportive leadership, rather than traumatic content as seen in commercial moderation~\cite{schopke-gonzalez_why_2024, roberts_behind_2019}. Such labor is further compounded by the expectation that volunteers manage complex interpersonal issues without the tools or backing typically available to professionals in similar roles.

The institutional and structural environment of online communities can significantly impacts motivation. For instance, \cite{halfaker_rise_2013} traced how Wikipedia’s increased automation and formalization in newcomer screening, initially for combating vandalism, unintentionally created barriers to participation. This shift likely contributed to a decline in active editors and plausibly intensified workloads for existing administrators. 

These studies underscore the multiple, intersecting factors affecting moderator motivation and well-being: invisible emotional labor, structural barriers for newcomers, and increasingly high coordination burden. Critically, however, existing research has not directly examined the experiences of Wikipedia administrators. Our study fills this important gap by surveying over 800 current and 2,000 potential admins, alongside interviews with 7 former and 5 current administrators. We thereby uncover the specific reasons admins may depart, what motivates current admins, and the challenges inherent to their role. Furthermore, by adapting prior survey methodologies to a new population ---Wikipedia administrators---, our work advances understandings of burnout in volunteer online moderation communities, specifically assessing its impact on Wikipedia administrators.

\section{Methodology}
\label{sec:methodology}

\subsection{Research approach}
This study employed a mixed-methods approach to investigate the recruitment, retention, and attrition of Wikipedia administrators. We combined quantitative analysis of publicly-available data from 284 Wikipedia language editions\footnote{Each language edition of Wikipedia is a separate independently-produced and governed project, not a direct translation of any other language version. A list of current Wikipedia language editions can be found at \url{https://en.wikipedia.org/wiki/List_of_Wikipedias}.}, with a focused examination of 21 larger Wikipedias. This was complemented by original surveys fielded in 6 language editions targeting current and potential administrators, and 12 interviews with current and former administrators across 5 language editions.

\subsection{Definitions}
We begin by defining the concepts, measurements, and populations used in this study.

\subsubsection{Defining administrator on Wikipedia across different language versions}
\label{sssec:definitions}
The basic assumption that  \textit{administrators do administrative work} is not straightforward to operationalize. The common definition of administrator on Wikipedia is generally  \textit{a user who is in the sysop (`system operator') user group}.\footnote{``User groups are groups that an editor can apply for in order to receive advanced tools to help them with editing the encyclopedia" `\url{https://en.wikipedia.org/wiki/Wikipedia:User_groups} } However, not all administrative rights are associated exclusively with the sysop user group, and therefore some members of non-sysop groups can perform administrative actions. For instance, on the Russian, Japanese, and Farsi Wikipedias, about a third of users performing administrator actions are not in the sysop user group\footnote{Examples of such user groups include \textit{eliminator}, \textit{closer}, and \textit{botadmin}; not all of these groups are present on all wikis, and even where they are, they do not need to have the same distribution of user permissions.}. This points to some degree of dispersal within these Wikipedias, where administrative responsibilities (and therefore, the user rights needed to take on those responsibilities) normally associated with sysops, are extended to other user groups. 


\subsubsection{Operationalizing definitions and measurements} \label{definitions}
Due to considerable variation  in how individual Wikipedia language editions disperse administrative rights, we introduce the following formal definitions in order to facilitate systematic between-Wikipedia comparisons: 
\begin{itemize}
\item \textit{Administrative action}: These actions can be related to content or user administrative tasks such as blocking/unblocking users or  protecting/deleting pages and are recorded in the MediaWiki logging table.\footnote{Every log action in MediaWiki is logged in the \href{https://www.mediawiki.org/wiki/Manual:Logging_table}{logging table}. Similar to how, \textit{e.g.}, the \href{https://www.mediawiki.org/wiki/Manual:Revision_table}{revision table} contains metadata for every edit done to a page within the wiki, the \href{https://www.mediawiki.org/wiki/Manual:Logging_table}{logging table} contains a record of other actions users perform on-wiki, such as blocking/unblocking users, deleting pages, setting page permissions, changing user rights, tagging, thanking, and uploading. Public log data are searchable at \href{https://www.mediawiki.org/wiki/Special:Log}{https://www.mediawiki.org/wiki/Special:Log} and fetchable via \href{https://www.mediawiki.org/wiki/API:Logevent}{https://www.mediawiki.org/wiki/API:Logevent}.} Specifically, any of the following recorded logged actions are considered administrative actions: \verb|delete| (delete page), \verb|event| (delete event), \verb|restore| (restore a page), \verb|revision| (change visibility), \verb|block|, \verb|reblock|, \verb|unblock|, \verb|protect|, \verb|modify|, \verb|unprotect|, \verb|rights| (change user group rights). 

\item \textit{Administrator} or \textit{`admin'}: Any user belonging to a user group with the ability to carry out one or more administrative actions.
\item \textit{Sysop} -- User in the sysop user group (a qualifying, but not necessarily required, condition of  administrator status); short for \textit{system operator.}
\item \textit{Active administrator}: Administrator who took $\geq1$ administrative action during a specified time period. A \textit{monthly active administrator} (MAA) is an administrator who took $\geq1$ administrative action within a specified calendar month
\item \textit{Former administrator}: User who previously held administrator rights on a given Wikipedia language edition, who no longer has those rights. For interview recruitment we narrow down to users who previously held administrator rights on one of English, Spanish, French, Indonesian, or Russian Wikipedia and no longer did by October 1, 2024 (the first day of interview recruitment), but remained active Wikipedia editors\footnote{A formal definition for `active editor' can be found at \url{https://meta.wikimedia.org/wiki/Research:Active_editor}} (made five or more content edits within the past month) or continued to attend broader Wikimedia Movement events\footnote{We prioritized contacting former administrators with some level of activity on Wikimedia projects as measured by their editing activity or participation in events. In practice we contacted administrators outside of the criteria as well when we did not receive sufficient responses from the prioritized group. However, we received zero responses from completely inactive former admins.} (\textit{e.g.}, virtual or in-person meet-ups). 
\item \textit{Potential administrator}: User who meets formal criteria for the Request for Adminship (RFA) process as determined by their Wikipedia language edition, but who is not currently an administrator.
\end{itemize}

\subsection{Case selection}
Case selection for each phase and method of analysis is described in this section and is summarized in Table~\ref{tab:data_sources}.

\begin{table}[!ht]
    \centering
        \caption{Case selection and scope by data source.}
    \scalebox{0.76}{
    \begin{tabular}{l|c|l}
    \toprule
     Data Sources & Wikipedias & Scope  \\
     \midrule
     Public data from the MediaWiki logging table & 284 & All current Wikipedia language editions which were ``live'' (\textit{i.e.}, hosted by the\\
     MediaWiki\_history table, user group table & & Wikimedia Foundation) by January 2018. \\
     \midrule
     All listed above \textbf{and} public content from: & 21 & All current Wikipedia language editions which were ``live'' by January 2018 \textbf{and} \\
     Wikipedia policy pages and discussions, admin & & had $\geq20$ monthly active admins (MAA) on October 1, 2024 \\
     pages and discussions, RfA discussion and votes & &  \\
     \midrule
     Surveys of current Wikipedia administrators & 6 & All admins (users in a user group with the ability to make changes to user \\
     & & blocks, page protection levels, page deletion status, or the rights of other users) \\
     & & on the English, Spanish, French, Indonesian, Polish,  and Russian Wikipedias \\
     \midrule
     Surveys of potential Wikipedia administrators & 6 & Users meeting the formal rules for RFA on the English, Spanish, French, \\
     & & Indonesian, Polish,  and Russian Wikipedias (see Table~\ref{tab:survey_combo_admins} for sampling rates) \\
     \midrule
     Interviews of current Wikipedia administrators & 5 & Random sample of current administrator survey respondents from English, Spanish, \\
     & & French, and Indonesian Wikipedias who volunteered to participate in interviews\\
     \midrule
     Interviews of former Wikipedia administrators & 5 & Former administrators on the English, Spanish, French, Indonesian, and Russian \\
     & & Wikipedias who are current active editors or attended recent Wikimedia \\
     & & community events \\
    \bottomrule
    \end{tabular}
    }
    \Description[Case selection and scope by data source]{Case selection and scope by data source}
    \label{tab:data_sources}
\end{table}

\subsubsection{Public data}
This study presents analysis of all 284 Wikipedia language editions that were live (\textit{i.e.}, hosted by the Wikimedia Foundation) as of January 2018\footnote{Another way of defining this subgroup would be all current Wikipedia language editions \textit{excluding} those not yet hosted outside of the Wikimedia Incubator by January 1, 2018, as shown at \url{https://incubator.wikimedia.org/wiki/Incubator:Site_creation_log}}, with a particular focus on the 21 Wikipedia language editions with 20 or more monthly active administrators as of January 2024, the last date for which we had access to the full year of data as of October 1, 2024 (See Table \ref{tab:survey_select} for the list of languages). We sometimes refer to these languages as ``shortlisted languages''. The threshold of 20 used for identifying the shortlisted languages is twice the mean number of MAA across all independently-administered\footnote{Wikipedia language editions with fewer than three monthly active administrators fall under the purview of global administrators.\cite{global_sysops_2025}} Wikipedia language editions. We consequently focus on language editions with robust administrator bodies and well-developed administrator candidacy policies. 
Wikipedias with lower (3-19) MAA numbers are likely to have different administrator needs, capabilities, and dynamics\footnote{As demonstrated by the existence of groups like the \href{https://meta.wikimedia.org/wiki/Small_Wiki_Monitoring_Team}{Small Wiki Monitoring Team}, as well as work published by the Wikimedia Foundation indicating that wikis with fewer admins (active or otherwise) face different administrator dynamics \cite{lo_content_2022, morgan_researchpatrolling_2019}} and research on these projects presents privacy risks due to the small number of MAA. 

In addition to compiling publicly available log data, we also studied those 21 Wikipedia language editions' current and archived policies related to administrator candidacy and tenure, discussions relating to the health of the administrator body, and any major proposed policy changes to any of the above, from 2018 to 2024.

\subsubsection{Survey data}
Survey data was collected from a subgroup of 6 Wikipedia language editions (English, Spanish, French, Indonesian, Polish, Russian). These language editions were selected to produce a ``diverse'' set of case studies \citep{seawright2008case} that would represent as much combined meaningful variation as possible in the number of MAA, admin activity levels (median actions per admin), RFA success rates, and admin candidacy guidelines among the 21 shortlisted Wikipedia language editions shown in Table~\ref{tab:survey_select}. 

\begin{table}[]
\centering
        \caption{Survey and interview case selection criteria.}
\resizebox{\columnwidth}{!}{%
\begin{tabular}{@{}lllllllll@{}}
\toprule
 & \multicolumn{1}{c}{} &  &  &  &  & \multicolumn{3}{c}{\textbf{Admin candidacy guidelines}} \\ \cmidrule(l){7-9} 
\multirow{-2}{*}{\textbf{Wiki}} & \multicolumn{1}{c}{\multirow{-2}{*}{\textbf{Admins}}} & \multirow{-2}{*}{\textbf{\begin{tabular}[c]{@{}l@{}}MAA \\ (2023*)\end{tabular}}} & \multirow{-2}{*}{\textbf{\begin{tabular}[c]{@{}l@{}}Median \\ actions/\\ admin\\ (2023*)\end{tabular}}} & \multirow{-2}{*}{\textbf{\begin{tabular}[c]{@{}l@{}}RFA \\ success\\ (2013-\\2023\textsuperscript{\textdagger})\end{tabular}}} & \multirow{-2}{*}{\textbf{Censorship}} & \textbf{\begin{tabular}[c]{@{}l@{}}Account age\\  (months)\end{tabular}} & \textbf{\begin{tabular}[c]{@{}l@{}} Article\\ edit \\count\end{tabular}} & \textbf{Other} \\ \\ \midrule
\rowcolor[HTML]{ECF4FF} 
\textbf{English (en)} & 911 & 398 & 19 & 50\% &  & 6 & None stated & \begin{tabular}[c]{@{}l@{}}Email recommended but not required; \\ part of extended-confirmed user group\end{tabular} \\  \midrule
\rowcolor[HTML]{ECF4FF} 
\textbf{Spanish (es)} & 60 & 44 & 314 & 28\% &  & 12 & None stated & Candidate must have email \\ \midrule
German (de) & 186 & 125 & 178 & 57\% &  & 24 & 1000 &  \\ \midrule
Japanese (ja) & 41 & 30 & 149 & — &  & 4 & 500 &  \\ \midrule
\rowcolor[HTML]{ECF4FF} 
\textbf{French (fr)} & 155 & 88 & 102 & 71\% &  & 12 & 3000 &  \\ \midrule
\rowcolor[HTML]{ECF4FF} 
\textbf{Russian (ru)} & 74 & 87 & 169 & 37\% & \begin{tabular}[c]{@{}l@{}}Current partial block; \\ previously blocked content\end{tabular} & 3 & 100 &  \\ \midrule
Chinese (zh) & 67 & 27 & 13 & 24\% & \begin{tabular}[c]{@{}l@{}}China currently blocking all content; \\ COI editing by state; prosecuted editors\end{tabular} & 12 & 3000 & \begin{tabular}[c]{@{}l@{}}Membership in patroller or rollbacker user groups; \\ not blocked in last year\end{tabular} \\ \midrule
Italian (it) & 120 & 109 & 302 & — &  & None stated & None stated & Member of autopatrolled group \\ \midrule
Portuguese (pt) & 52 & 70 & 58 & — &  & 12 & 300 &  \\ \midrule
Persian (fa) & 35 & 31 & — & 36\% & \begin{tabular}[c]{@{}l@{}}Current partial block in Iran; \\ previously blocked content\end{tabular} & 6 & 1000 &  \\ \midrule
\rowcolor[HTML]{ECF4FF} 
\textbf{Polish (pl)} & 100 & 72 & 128 & 51\% &  & 3 & 500 & Candidate must have email \\ \midrule
\rowcolor[HTML]{ECF4FF} 
\textbf{Indonesian (id)} & 44 & 30 & 127 & 60\% &  & 3 & 500 & Email; user page > 500bytes; no blocks in past 6mo \\ \midrule
Dutch (nl) & 34 & 31 & 111 & 68\% &  & 6 & 1000 & Candidate must have email \\ \midrule
Ukrainian (uk) & 47 & 38 & 150 & 50\% &  & 6 & 2000 & \begin{tabular}[c]{@{}l@{}}At least 200 additional edits in service namespace \\ (Help, Template, Module, Wikipedia)\end{tabular} \\ \midrule
Hebrew (he) & 29 & 30 & 970 & — &  & 9 & 2000 &  \\ \midrule
Czech (cs) & 31 & 25 & 183 & — &  & 6 & 250 & Candidate must have email \\ \midrule
Swedish (sv) & 66 & 51 & 128 & — &  & None stated & None stated &  \\ \midrule
Finnish (fi) & 33 & 20 & 34 & — &  & “A few” months & None stated &  \\ \midrule
Norwegian Bokmål (no) & 44 & 32 & 106 & — &  & 4 & 1000 & Has email and user page \\ \midrule
Catalan (ca) & 29 & 20 & 174 & — &  & 9 & 1000 & Requirements presented “as guidelines” \\ \midrule
\multicolumn{9}{l}{"—" denotes metrics for which data could not be found.} \\ 
\multicolumn{9}{l}{*2023 was used as a reference year as it was the most-recent full year when the surveys were conducted in late 2024.} \\ \multicolumn{9}{l}{\textsuperscript{\textdagger}Because RFA success rates are highly variable year-to-year, we averaged them over a ten-year period. We begin this period in 2013 because the RFA process was fundamentally different in} \\
\multicolumn{9}{l}{Wikipedia's first decade and featured dramatically higher success rates.} \\ \bottomrule
\end{tabular}%
    \Description[Interview and survey case selection]{Interview and survey case selection}
    \label{tab:survey_select}
}
\end{table}

\subsubsection{Interviews}
Semi-structured interviews were conducted with current and former administrators on English, Spanish, French, and Indonesian, and Russian Wikipedia.

\subsection{Data collection}
\subsubsection{Public data collection}
We collected publicly-available administrative data from the aforementioned MediaWiki login table. This includes:
    \begin{itemize}
        \item \textit{Admin inflow}: a measure of the total number of users who gained administrative rights in a given time period 
        \item \textit{Admin outflow}: a measure of the total number of users who lost administrative rights in a a given time period (a measure of net administrator recruitment)
        \item \textit{Net admin recruitment}: inflow minus outflow during a given time period; a negative number indicates higher attrition than recruitment  
    \item \textit{Monthly active administrators (MAA)}: a measure of the total number of administrators who performed at least one administrative action in a given month
    \item \textit{Monthly administrative actions}: the total number of blocks/unblocks, page deletions/restorations, page protection changes, and user rights changes in a given month
\end{itemize}

Public data on Wikipedia administrative activity was collected for the period from January 2018 to June 2025, the last day we re-ran the quantitative analyses. A January 2018 start date was used due to known data quality issues\footnote{\textit{E.g.}, Prior to August 2016, no deletion log event was generated when a page was moved on top of a redirect; instead it created an \href{https://phabricator.wikimedia.org/T106119}{empty row in the logging data}. However, when the log event \href{https://phabricator.wikimedia.org/T106119}{was added} for this action, it used the same \texttt{log\_action} as regular deletions: \texttt{delete}. This resulted in these particular actions being indistinguishable from page deletions (the latter of which are administrative actions, the former of which are not) in the logging data. In November 2016, the \texttt{delete\_redir} action was created to allow the two types of events to be distinguished. So \texttt{delete} counts in the logging data are overinflated between August and November 2016. (For details, see \href{https://phabricator.wikimedia.org/T145991}{Phabricator T145991} and \href{https://gerrit.wikimedia.org/g/mediawiki/core/+/71e336981b5db53c9db614c18ee798863c3ded9c}{Gerrit bug fix}).}, Wikipedia-edition-specific policy changes\footnote{Per \href{https://phabricator.wikimedia.org/T341879}{T341879}, there are drops in admin counts when certain wikis adopted policies making inactive admins lose their rights; see \href{https://en.wikipedia.org/wiki/Wikipedia:Desysoppings_by_month\#Notes:~:text=than\%2048\%20desysoppings-,Notes,-\%5E\%20Figures\%20don\%27t}{Notes} for details.}, and alignment with the years included in WMF's public \href{https://meta.wikimedia.org/wiki/Movement_Insights/Wiki_comparison}{wiki comparison} data to facilitate comparisons.


\subsubsection{Survey data collection}
We fielded a series of surveys of current and potential administrators on six Wikipedia language editions from October 23 and November 25, 2024\footnote{Later fielding dates for Polish and Russian Wikipedia surveys (See Table~\ref{tab:survey_combo_admins}) reflect logistical needs (availability of completed translations) rather than an explicit research design.}. Surveys were programmed and administered via Limesurvey and were distributed via MassMessage\footnote{See \url{https://meta.wikimedia.org/wiki/MassMessage} for documentation of this tool and its capabilities}, a tool that allows users with specific permissions on Wikimedia projects to contact any list of editors on a given project through their email. All users with administrator rights on one of the target Wikipedias as of October 1 2024 were considered eligible current administrators. All editors who as of October 1, 2024, met their language edition's formal eligibility criteria for requesting adminship (RFA), but who had not yet done so were considered eligible potential administrators. Prospective survey respondents were provided with a privacy statement outlining data retention and publication policies. All survey responses have been anonymized.

As shown in Table~\ref{tab:survey_combo_admins}, due to the small total population of current administrators, we contacted the entire eligible populations of current administrators (100\% sampling rate) in each surveyed language edition. A total of 1285 current administrators were invited to take \textit{``a survey of Wikipedians to better understand what draws administrators to contribute to Wikipedia and what affects administrator retention''} via MassMessage, of whom 805 (62.6\%) provided at least a partial response.

\begin{table}[ht]
    \caption{Fielding dates, population, number of contacts and responses for current and potential administrator surveys on six target language editions of Wikipedia. Fielding dates can vary depending on the availability of translations and response rate but were capped at November 25.}
    \centering
    \scalebox{0.8}{
        \begin{tabular}{ l | l | c | c | c | c | c }
        \toprule
        Sample & Fielding Dates & Eligible users & Contacts & Sampling rate & Total responses & Response rate \\
        \midrule
        English Current & October 23--November 25, 2024 & 838 & 838 & 100\% & 508 & 60.6\% \\
        French Current & October 23--November 25, 2024 & 145 & 145 & 100\% & 98 & 67.6\% \\
        Spanish Current & October 24--November 25, 2024 & 56 & 56 & 100\% & 41 & 73.2\% \\
        Indonesian Current & October 24--November 25, 2024 & 47 & 47 & 100\% &  27 & 57.4\% \\
        Polish Current & November 8--November 18, 2024 & 97 & 97 & 100\% & 66 & 68.0\% \\
        Russian Current & November 8--November 25, 2024 & 102 & 102 & 100\% &  65 & 63.7\% \\
        \midrule
        Total Current & October 23--November 25, 2024 & 1285 & 1285 & 100\% & 805 & 62.6\% \\
        \midrule
        English Potential & October 23--November 25, 2024 & 4884 & 1650 & 33.78\% & 779 & 47.2\% \\
        French Potential & October 23--November 25, 2024 & 195 & 195 & 100\% & 134 & 68.7\% \\
        Spanish Potential & October 23--November 25, 2024 & 13694 & 1780 & 13.0\% & 680 & 38.2\% \\
        Indonesian Potential & October 23--November 25, 2024 & 24 & 24 & 100\% & 18 & 75.0\% \\
        Polish Potential & November 8--November 25, 2024 & 2544 & 1530 & 60.1\% & 265 & 17.3\% \\
        Russian Potential & November 8--November 25, 2024 & 1736 & 1736 & 100\% & 368 & 21.2\% \\
        \midrule
        Total Potential & October 23--November 25, 2024 & 23077 & 6915 & 30.0\% & 2244 & 32.5\% \\
        \bottomrule
        \end{tabular}
    }
    \Description[Sampling, population and responses for potential administrator surveys]{Fielding dates, population, sampling rate and responses for current and potential administrator surveys on six target language editions of Wikipedia. }
    \label{tab:survey_combo_admins}
\end{table}

As summarized in Table~\ref{tab:survey_combo_admins}, the population of potential administrators can vary considerably by Wikipedia language edition due to variation in the formal RFA eligibility criteria. Thus, for larger potential administrator populations (>1500), we chose to randomly sample from the eligible population. Sampling rates were set to achieve a response target of approximately 500, based on an estimated response rate of 33\%\footnote{This was based on extrapolations from publicly-reported response rates to the Wikimedia Foundation 2024 Community Insights survey \cite{andic_research_ci_2024}}. However, response rates varied substantially by surveyed language edition with comparatively low response rates among potential administrators on Polish and Russian Wikipedia.

Whenever possible, survey questions were adapted from previously-published surveys of Wikipedia users (\textit{e.g.,} \cite{andic_research_ci_2024,liou_research_grs_2024,matias_2020} and well-validated government and academic surveys (\textit{e.g.,} \cite{uscensus_acs_2024,ESS_2021_Round_10}). Questions were evaluated for potential cross-cultural variation in social desirability (\textit{e.g.}, gender identity) as well as for concept equivalence (\textit{e.g.}, educational attainment levels) \cite{harkness_2016}. Because prior research has established high levels of educational attainment overall among Wikipedia editors \cite{andic_research_ci_2024}, questions were written in a more formal tone to facilitate translation and construct equivalence\footnote{Specifically, we sought to avoid colloquialisms and idiomatic expressions that can be difficult to translate between languages and cultural contexts \cite{salazar_2022}.}, both between survey language versions and within survey language versions (\textit{e.g.}, for respondents to the Spanish language survey from Spain and Latin America). Broadly speaking, we adopted a team translation approach \cite{mohler_2016}. Survey scripts and MassMessage invitations were prepared for translation using text templates allowing translators to view each survey question and answer option in context and to provide line-by-line translation from English. All surveys contained the same content regardless of language (All Survey Questions Translated). Paid translators were recruited from a pool of experienced Wikipedia contributors with simultaneous expertise in translation and as an active editor within the relevant Wikipedia language edition. Each translation was prepared by a single translator and no translator prepared more than one translation.

Translated survey scripts and MassMessage invitations were then validated via a multi-step process. First, translations were back-translated to English using machine translation (DeepL) as an initial assessment for alignment with the source materials. Potential issues were communicated to translators and resolved through collaboration between the survey author and individual translators. Second, a second group of reviewers with relevant language proficiency and Wikipedia domain knowledge separately reviewed translated survey scripts and MassMessage invitations (using the text template described above) as well as programmed surveys for consistency and validity\footnote{Pre-testing the survey scripts with ``real'' respondents was not feasible due to the small population sizes of current (and potential) Wikipedia administrators}. 

Survey respondents were screened by age; survey sessions for those under 18 were terminated. The current administrators' surveys were designed to take about 15-20 minutes to complete and contained modules covering the following general topics:
\begin{itemize}
    \item Screeners for age, sysop membership, and administrative actions taken in the past 30 days\footnote{Only the age screener was used to terminate survey sessions.}
    \item Experience as a contributor to Wikimedia projects prior to adminship (\textit{e.g.} years of experience as an editor, experience performing non-sysop moderation activity) 
    \item Experience with the RFA process (\textit{e.g.}, activities performed in anticipation of RFA process, assessment of personal RFA experience, assessment of RFA process for current adminship candidates)
    \item Day-to-day experience as a Wikipedia administrator (\textit{e.g.,} burnout assessment adapted from Matias et al. (2020) \cite{matias_2020}, motivations for work as an administrator, time spent on administrator work)
    \item Pain points or challenges experienced as a Wikipedia administrator (\textit{e.g.}, experience of conflict, consideration of quitting / reducing hours
    \item Demographics (\textit{e.g.}, gender, age, educational attainment)
    \item Opt-in opportunity for interview research
\end{itemize}
The potential administrators' surveys were designed to take 10-15 minutes to complete and covered the following general topics:
\begin{itemize}
    \item Age screener
    \item Experience as a contributor to Wikimedia projects
    \item Adminship familiarity and interest  (\textit{e.g.,} familiarity with the RFA process and the work of Wikipedia administrators, interest in adminship, reasons for interest or lack of interest)
    \item Demographics
\end{itemize}
Given the number of survey questions, we opted for sharing the complete list of survey questions in an anonymized way, and not in Appendix for the review process: \url{https://osf.io/j2hcw/?view_only=f9b222b6ad1e4fad9810bc3c0ed6b283}

\subsubsection{Interview data collection}
We conducted semi-structured interviews (Approximately 60 minutes each) with 5 current and 7 former admins (Wikipedia editors who had previously held administrator rights, but no longer did by October 1, 2024) from English, Spanish, French, Indonesian and Russian Wikipedia. All interviewees were provided with a privacy statement outlining data retention and publication policies. All interviews were anonymized by default. A sample of interview questions can be found in Table~\ref{tab:interview_questions} in Appendix. Interviews began by asking participants to describe their involvement with Wikipedia, when they first became an administrator, how long they had been an administrator, and when they lost administrator privileges (where relevant). The interviews then continued to ask about their experiences as an RFA candidate, their experiences of adminship and motivations for remaining in the role, and demotivating factors during their tenure as an administrator.

Current administrator interviewees were recruited from among survey participants who volunteered for interviews. Volunteers were randomly-selected to be contacted for interviews. Qualifying former administrators (see Section~\ref{sssec:definitions}) were identified as potential interviewees. Former administrators who left on bad terms, such as those blocked by their Wikipedia editions or the Wikimedia Foundation, were not contacted to mitigate the potential harm that it may cause to individuals or the communities.

Due to repeated censorship and attacks on Russian Wikipedia by the Russian government \cite{Reuters2022RussiaPunishWikimedia,Vice2022RussiaThreatensBlockWikipedia}, we only contacted specific former Russian Wikipedia administrators who were pre-screened for their location to assure they are outside of the Russian Federation for their safety. All potential interviewees were then contacted via email.

All interviews were conducted using Google Meet video-conferencing software, and audio-video recordings and transcripts were created for analysis. Interviewees were offered monetary incentives for completed sessions. Interviews were conducted by a single interviewer in the participant's preferred language, with interpreters available on request. Four individual interviewers (all experienced qualitative researchers) contributed to this research.

Interview analysis, conducted by the interviewers, was done through open coding \cite{saldana_coding_2021} to describe emerging themes through the transcripts and recordings, organized hierarchically under our main discussion categories (``becoming an administrator'', ``experiences and motivations as an admin'', ``demotivating factors'' and ``reasons for quitting''). Analysis involved coding of both responses to direct questions such as, ``Why did you seek out admin rights?'' and ``What caused you to lose your admin rights?'' as well as responses to more general prompts on related topics, such as whether or not they would advise any editor to seek out adminship. Codes used include descriptive ones meant to capture events such as ``harassment from editors'', ``policy disagreement'', or ``work-life balance'', as well as emotional ones meant to capture subjective experiences such as ``frustration'', ``exhaustion'', ``caution'', or ``confusion''. Codes were initially developed by one interviewer based on review of roughly half of the completed interviews and iteratively refined through a collaborative process with the other interviewers. Discrepancies in coding were reconciled by a consensus process involving all coders.

\section{Results}
\label{sec:results}

\subsection{Administrator trends: Stability or growth in some Wikipedias; decline in others.}
\label{sec:results-trends}
We analyzed Wikipedia administrator logs across 284 Wikipedia languages.  

We find that the number of average monthly Wikipedia administrators is steady or growing in the majority of Wikipedia languages. We computed the net change in average monthly active admins between 2018 and 2025 for the 284 Wikipedias and established that roughly half of the 284 Wikipedias (51\%) show a net increase in active admins since 2018, and roughly a third (33\%) show no net change since. 

\begin{figure}[!ht]
    \centering
    \includegraphics[width=1\linewidth]{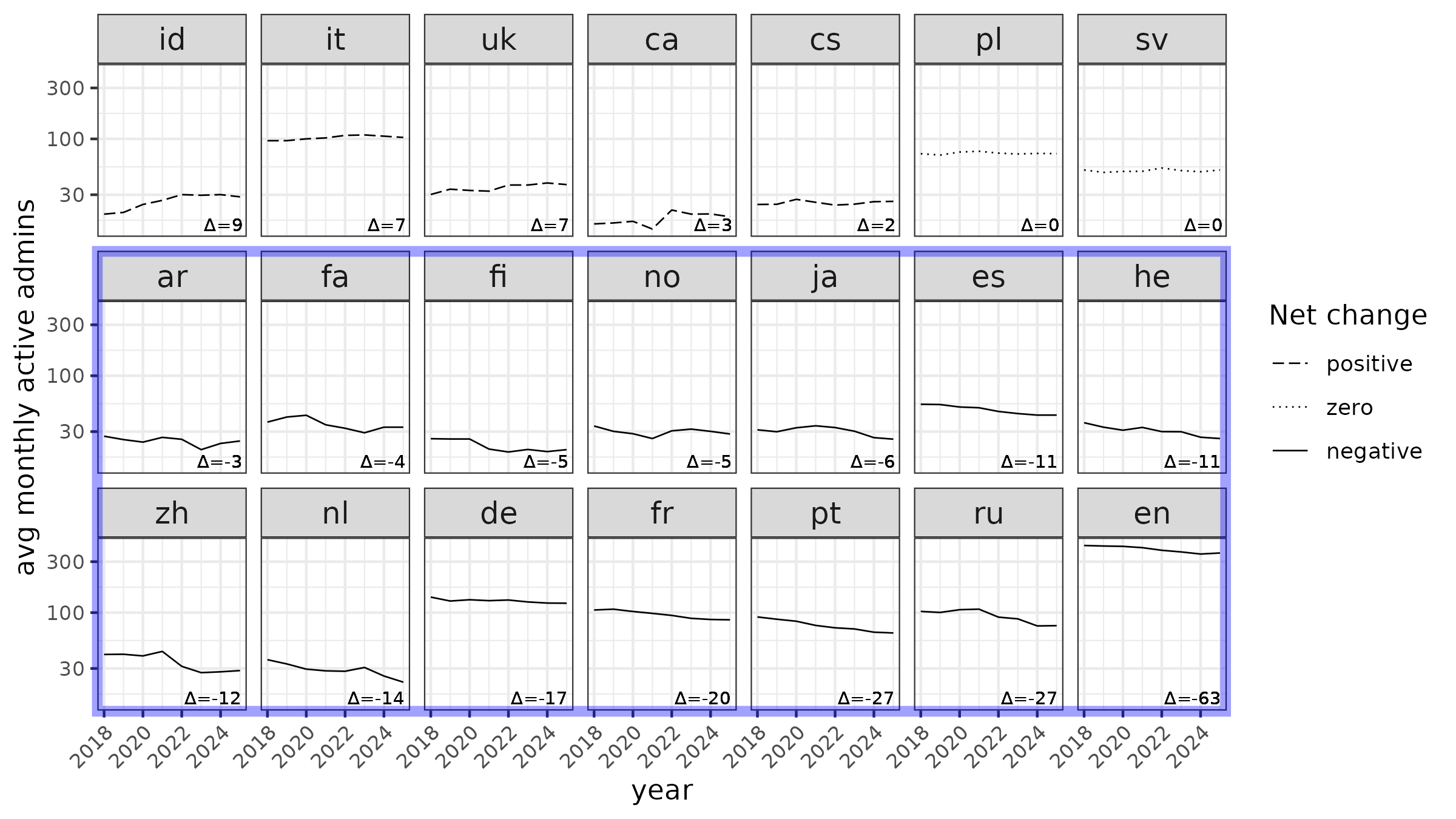}
    \caption{Per year average monthly active admins for the 21 shortlisted Wikipedias for years 2018 to 2025, ordered by net change ($\Delta$). Wikipedias with negative change are highlighted by a blue box. We plot yearly timepoints (\textit{i.e.,} the mean of monthly active admins across the calendar year) to avoid the effects of seasonality \cite{mittermeier2019season}. Specifically, we utilize means rather than raw monthly values--which vary due to both seasonality and idiosyncratic current events--that could give the impression of a decline when it is actually due to regression towards-the-(yearly) mean \cite{barnett2005regression}.}
    \label{fig:active-admins-over-time}
    \Description[line plots with avg monthly active admins for 21 wikis]{Average monthly active administrators for 21 Wikipedia language editions between 2018 and 2025 are shown. The graphs are ordered by net change from most positive to most negative. The top row shows languages with positive or zero change, the middle row shows those with a small negative change, and the bottom row shows those with the most significant decline. The y-axis uses a log scale, to better visualize the data which span a large range of values.}
\end{figure}

However, the number of average monthly Wikipedia administrators is declining in the majority of the 21 shortlisted Wikipedias with higher number of monthly active administrators  (see Table~\ref{tab:survey_select}  for list of shortlisted Wikipedias; see Table~\ref{tab:wiki-activity-2018-2024}  for additional activity-related metrics for these 21 Wikipedias). Roughly two-thirds (14 Wikipedias, 67\%) show a net decline in average monthly active admins since 2018, which amounts to 15\% of Wikipedia languages studied. These Wikipedias (in the order of increasing severity of decline as measured by net change in average monthly administrators) are: Arabic (ar), Farsi (fa), Finnish (fi), Norwegian (no), Japanese (ja), Spanish (es), Hebrew (he), Chinese (zh), Dutch (nl), German (de), French (fr), Portuguese (pt), Russian (ru) and English (en) Wikipedias (See Figure~\ref{fig:active-admins-over-time}).

These findings are significant as they demonstrate that while administrator activity is stable or increasing across the majority of Wikipedia languages, the majority of the largest Wikipedias by active administrator count are experiencing declines in active administrators.\footnote{Because time series data are prone to natural variation, we implemented multiple procedures to look for and mitigate potential effects of regression towards the mean. We ran linear models for each Wikipedia edition, with month as the predictor variable and number of active admins as the outcome variable. Results of these models indicate that 14 of the 21 wiki have negative slopes—which is the same finding we found in the results of the analysis visualized in Figure~\ref{fig:active-admins-over-time}. We then reran the linear models using a sliding 6-year window of time (starting with January 2018 to January 2024, and ending with January 2019 to January 2025) to examine the model outputs using multiple initial sample points. In the Appendix, Figure~\ref{fig:regression-slope-comparisons} shows the slope values from each model output; comparison of these slopes shows minimal change in the regression slopes of the Wikipedias regardless of initial sample points.} This contrasting trend underscores a potential shift in administrative dynamics between smaller and larger language communities on Wikipedia. For the remainder of this research we focus on identifying the underlying drivers, as well as challenges and opportunities for Wikipedia communities to change the trends.



\subsection{Insufficient recruitment: the primary driver of decline}
\label{sec:results-recruitment}
Our analysis indicates that the decline in administrator numbers across the 14 highly active Wikipedias is primarily driven by insufficient recruitment of new administrators, rather than an unusual increase in attrition.

First, we examined administrator departures (through removed admin rights). Data show annual administrator attrition in these Wikipedias remains remarkably steady, with few significant spikes in annual departures. Crucially, when moderate increases in departures do occur, they are often accompanied by increases in recruitment, suggesting a relatively stable balance in the rate of attrition. This stability is further supported by survey data: 75.6\% of current Wikipedia administrators agree (34.8\% strongly agree) that they expect to be administrators two years from now. Similarly, when asked how often they have considered ``\textit{quitting your role as...a Wikipedia administrator}'' 86.9\% report ``rarely'' or ``never'' considering quitting their administrator duties (67.8\% say ``never''). These findings strongly suggest that the observed decline is not due to administrators leaving at an accelerated rate.

\begin{figure}[!ht]
\centering
\includegraphics[width=1\linewidth]{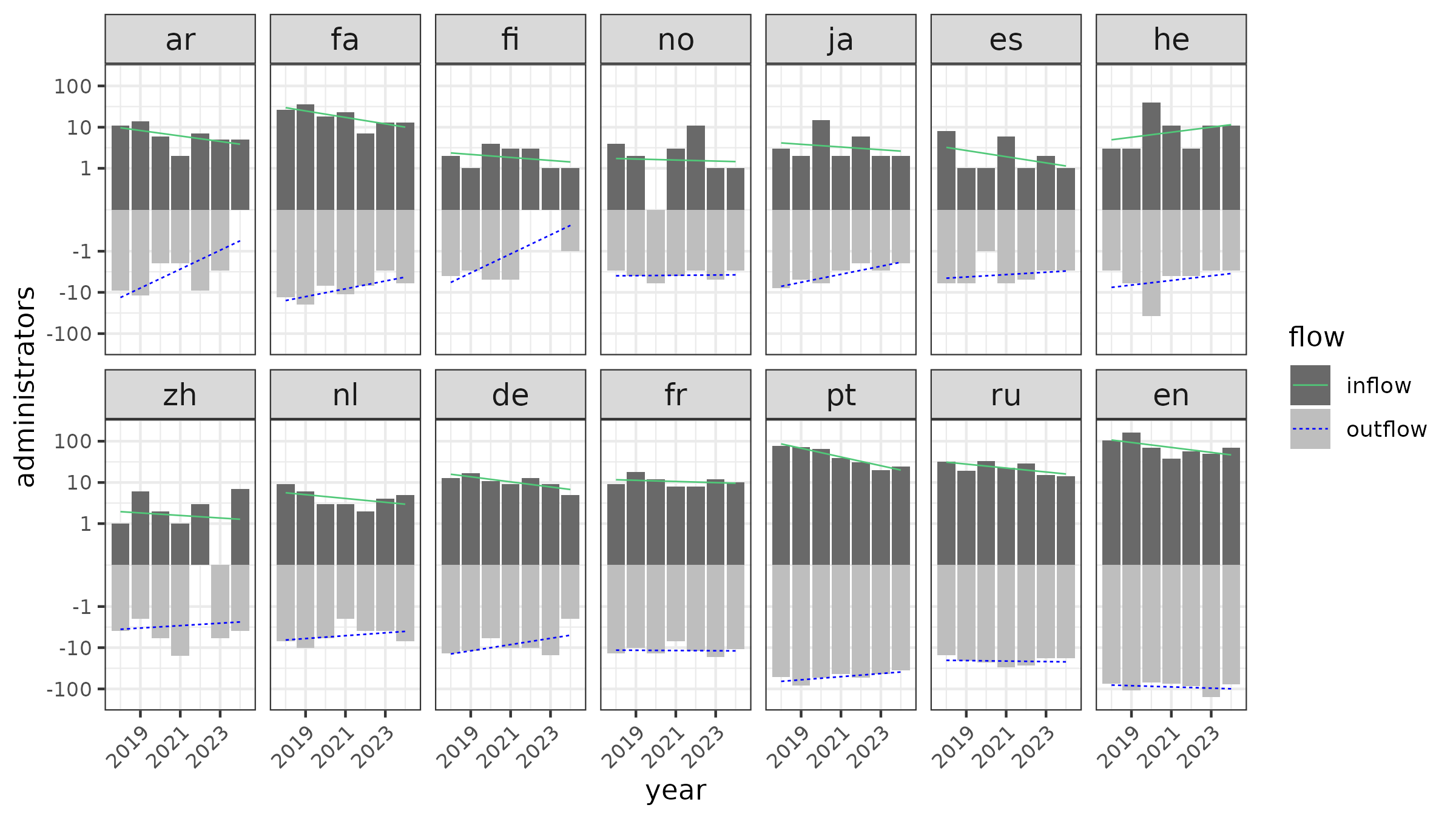}
\caption{Yearly admin inflow (admin right granted; recruited) and outflow (admin right removed; departed) across the 14 very active Wikipedias with an overall decline between 2018 and 2024 (See Figure~ \ref{fig:active-admins-over-time}). Trend lines are added to assist with visualizing the change in inflow and outflow across multiple years.}
\Description{This grid of bar plots illustrates the yearly inflow and outflow of administrators for 14 different Wikipedias from 2018 to 2024. The green bars show inflow (users gaining admin rights), and the blue bars show outflow (users losing rights). The plots generally show that outflow has been greater than inflow, leading to a negative net change for most language editions. The y-axis uses a log scale, to better visualize the data which span a large range of values because of larger numbers from English Wikipedia.}
\label{fig:inflow-outflow}
\end{figure}

However, this steady attrition rate becomes problematic when set against a persistent deficit in recruitment. Our analysis of annual inflow (newly granted admin rights) and outflow (removed admin rights) clearly demonstrates that for the majority of these Wikipedias (10 out of 14), the number of newly-recruited administrators is consistently insufficient to compensate for departures (Figure~\ref{fig:inflow-outflow}). This quantitative evidence—showing inflow persistently lower than outflow—is the primary mechanism driving the net decline in admin numbers. This conclusion aligns with current administrators' own assessment, with recruitment of new administrators frequently cited as an ``essential priority'' in survey responses (along with conflict resolution resources).

While insufficient recruitment is more the primary driver of the overall decline trend, understanding the reasons behind individual administrator departures remains critical for long-term retention and recruitment strategies. Our mixed-method analysis reveals several factors contributing to voluntary departures, none of which, however, indicate a systemic or unusual surge in attrition. Interviews with former administrators consistently cited three main categories: changing life circumstances leading to a loss of free time; interpersonal conflict from within the editor or administrator body; and burnout from the perceived pressures of ``admin work'' over personally enjoyable tasks. For instance, one former French Wikipedia administrator noted, \textit{"For months, I thought I was not working enough [as an admin] on Wikipedia and I was not a good administrator... It was me who thought it was not responsible to stay."} These qualitative insights are corroborated by survey data, which highlight challenges such as balancing admin roles with outside responsibilities (35.6\% agree) and the near-universal experience of interpersonal conflict (95.8\% report experiencing it), often involving abuse or harassment (68.7\%).

Having established insufficient recruitment as the primary driver of decline, the next logical question is why communities are struggling to recruit sufficient administrators.

\subsection{The challenges of recruiting Wikipedia administrators}
\label{sec:results-challenges}
Our multi-method analysis across 21 shortlisted Wikipedias investigated the challenges in administrator recruitment. Our data sources included a review of current and archived Wikipedia policies, historical administrator candidacy outcomes (from 2015 to 2024), and community discussions. We also analyzed over 3000 survey responses from current and potential administrators (805 and 2244 respectively), complemented by semi-structured interviews with 5 current and 7 former administrators. This comprehensive approach allowed us to identify prominent barriers to recruiting new administrators, including limited awareness of the role, ambiguous and demanding candidacy processes, and low overall interest among potential candidates. We also consider the current administrators' perspective on these challenges, along with their motivations, which represent a crucial strength.


\subsubsection{Awareness of adminship is limited among potential administrators.} 
\begin{figure}[!ht]
  \includegraphics[width=\textwidth]{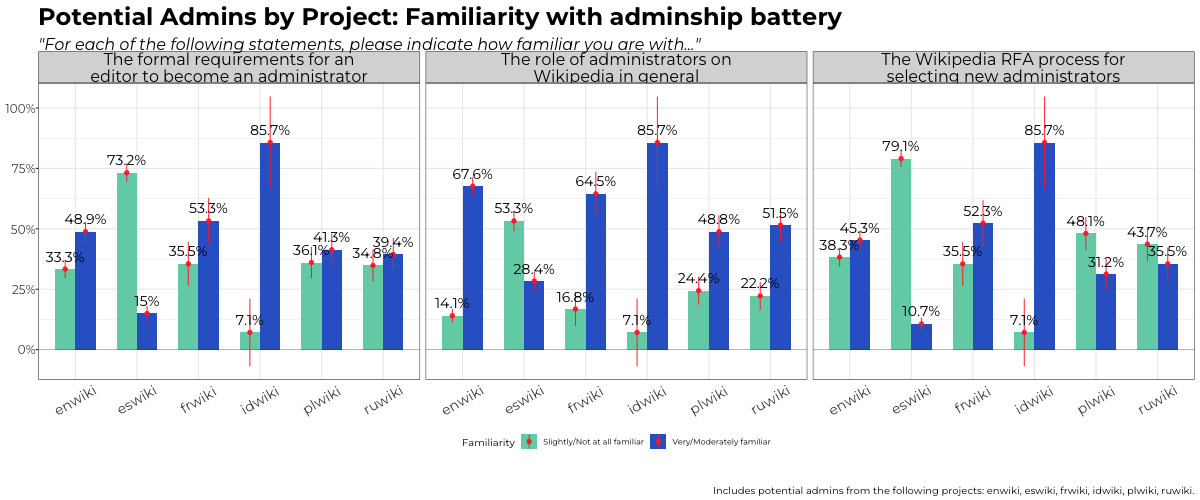}
  \caption{How familiar potential Wikipedia administrators (those that meet the formal requirements for adminship candidacy on their respective projects) are with different aspects of Wikipedia adminship. From a survey of potential English, Spanish, French, Indonesian, Polish, and Russian Wikipedia administrators. Sample proportion estimates are shown with 95\%\ confidence intervals.}
  \Description{Barchart showing familiarity with different aspects of Wikipedia adminship by potential Wikipedia admins}
  \label{fig:admin-familiarity}
\end{figure}

Our survey revealed that awareness of the administrator role and the RFA process for selecting new admins is limited among potential candidates and varies considerably by language edition (See Figure~\ref{fig:admin-familiarity}). While more than half of potential administrators in 4 out of 6 Wikipedias surveyed report being at least ``moderately familiar'' with the administrator role (English, 67.6\%; French, 64.5\%; Indonesian, 85.7\%; Russian, 51.5\%) familiarity drops substantially in the majority of the surveyed projects when it comes to the RFA process and the formal requirements to become an administrator. 

Notably, the proportion of potential Wikipedia administrators saying they are only ``slightly'' or ``not at all familiar'' with the RFA process ranges from substantial minorities on  English (38.3\%) and French (35.5\%) Wikipedia to pluralities on Polish (48.1\%) and Russian (43.7\%) Wikipedia. Potential administrators on Spanish Wikipedia report much lower levels of familiarity (79.1\% ``slightly'' or ``not at all familiar'') while those on Indonesian Wikipedia report much higher levels (just 7.1\%). We observe the same patterns across language editions for familiarity with the formal requirements for adminship.\footnote{We note that in all three categories, potential Indonesian Wikipedia administrators report distinctively high levels of familiarity (although this is based on a sample of 14 respondents) while potential Spanish Wikipedia administrators reported distinctively low levels of familiarity overall. Determining why such patterns exist is outside of the scope of this research. However, we note that the significant differences may reflect the very low ratio of potential admins:current admins on Indonesian Wikipedia (24:47) and conversely, the very high corresponding ratio on Spanish Wikipedia (13,694:56). In addition, Spanish Wikipedia has a distinctively small and active admin population and low RFA success rate (See Table~\ref{tab:survey_select}), particularly in relation to the overall size of the project \cite{movement_insights_wiki_2024}.}


\subsubsection{Interest in adminship is generally low.}
Interest in serving as a Wikipedia administrator is low across most language editions surveyed. Interest is particularly low among potential admins in English and Polish Wikipedia where 62.7\% and 58.1\% of potential administrators, respectively, say they are only ``slightly'' or ``not at all'' interested in adminship. In contrast, potential administrators on Spanish (49.8\%), Russian (48.9\%), and French (47.6\%) Wikipedia show relatively higher interest, with about half of potential candidates indicating that they are at least ``somewhat'' interested. 

 \begin{figure}[!ht]
  \includegraphics[width=\textwidth]{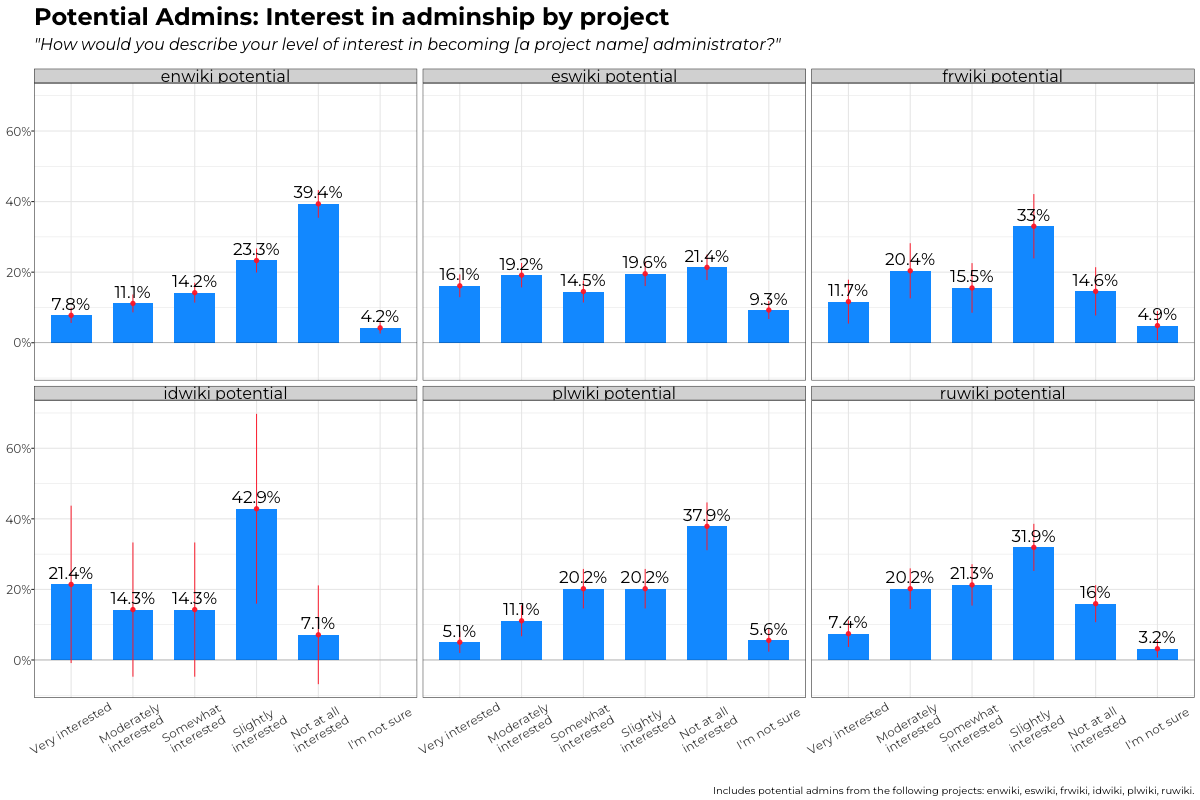}
  \caption{Self-reported prospective interest in adminship by potential Wikipedia administrators. From a survey of potential English, Spanish, French, Indonesian, Polish, and Russian Wikipedia administrators. Sample proportion estimates are shown with 95\%\ confidence intervals.}
  \Description{Barchart showing interest levels in Wikipedia adminship by potential Wikipedia admins}
  \label{fig:admin-interest}
\end{figure}

The most common reasons given for lack of interest in adminship, in the order of frequency reported\footnote{Because these results are generally stable across language editions, we present pooled estimates here for ease of exposition.}, are: 1) desire to focus on editing rather than administrative tasks (66.2\%), 2) perception that admin work is too time-consuming (54.7\%), and 3) concerns about conflict with other editors (34.7\%), particularly on French Wikipedia (55.1\%). Notably, a desire to focus on editing and concern that admin work is too time-consuming were among the top three reasons cited for lack of interest among potential admins in all six languages surveyed. 
Some survey respondents described this by saying that becoming an admin would require them to ``self-censor'' or that they would feel ``an obligation to contribute''.
Conversely, few respondents cited any aspect of the RFA process (meeting formal requirements, navigating the process, or perceptions of the process as unfair) as a reason for their lack of interest.
However, survey respondents who did cite the RFA process as a reason for lack of interest were emphatic about it, calling it ``stressful'', ``toxic'' and ``a vehicle for degrading some applicants.'' 

\begin{quote}
    \textit{``One has to be a saint... to get past the nit-pickers at RFA.''} -- potential English Wikipedia administrator
\end{quote}

Of the potential candidates who indicated they were interested in seeking administrator status, altruistic motivations were the most common, with respondents saying that they wanted to ``[help] underrepresented communities find a voice'', or to ``promote the democratization of knowledge''. Candidates interested in adminship also saw it as a way to ``avoid the bureaucracy of having to formally request others complete simple tasks'', or saying that lacking administrators who were also subject matter experts in their areas of interest ``hinders my editing activities''.

\subsubsection{Ambiguous and varied requirements as well as a demanding, stressful RFA process.}
Our review of Wikipedias' policies related to administrator candidacy and term limits found that requirements for admin candidacy vary widely by Wikipedia language edition, and are often a mix of formal and informal criteria. Furthermore, the informal requirements can vary even among individual administrators.  

The formal requirements can vary significantly from project to project. For example, we find that account age guidelines range from no minimum, to ``a few months'', to a full 2 years. Factoring in both formal and informal guidelines, requirements for minimum article edit counts for administrator candidacy similarly range from no minimum, to 100, and even 10,000. Language editions may also list additional guidelines, such as membership in a user group (\textit{e.g.}, \textit{patroller}, \textit{rollbacker}, \textit{autopatrolled}), presence of an email address, or evidence of participation in anti-vandalism work. 

Informal norms can be even more influential than formal criteria. For instance, on English and Indonesian Wikipedia, interviewees reported that sponsorship by an existing administrator is an informal but critical expectation, with self-nomination sometimes seen as a misunderstanding of community norms and grounds for disqualification. A repeated theme in our interviews was that while formal guidelines serve to filter out obviously unqualified candidates, actual decisions are often based on broader, unwritten standards.

These varying and often unwritten requirements contribute significantly to the perceived difficulty and stress of the Request for Adminship (RFA) process itself, which potential, current, and former administrators consistently describe as demanding.

\begin{figure}[!ht]
  \includegraphics[width=\textwidth]{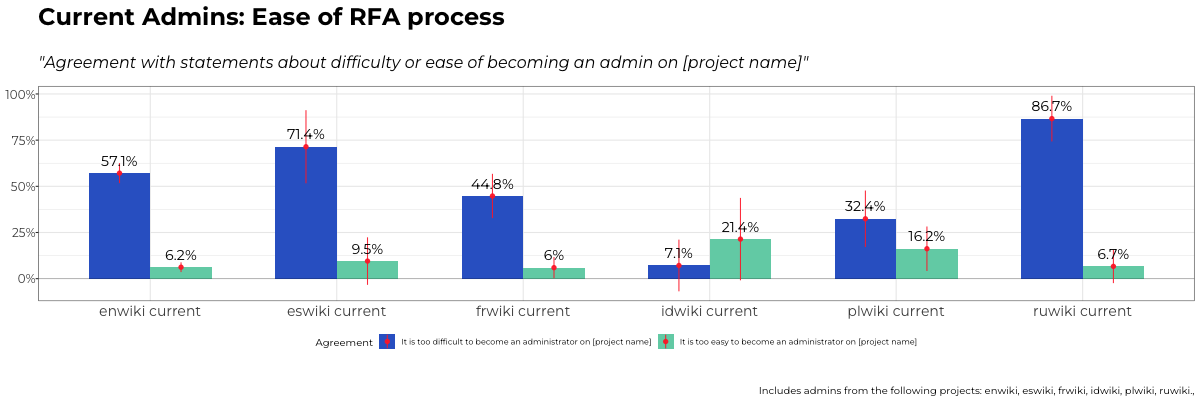}
  \caption{Current administrators' agreement with statements about the difficulty of becoming an admin on their project.}
  \Description{Barchart showing attitudes on how easy it is to become an admin.}
  \label{fig:rfa_ease}
\end{figure}

\begin{figure}[!ht]
  \includegraphics[width=\textwidth]{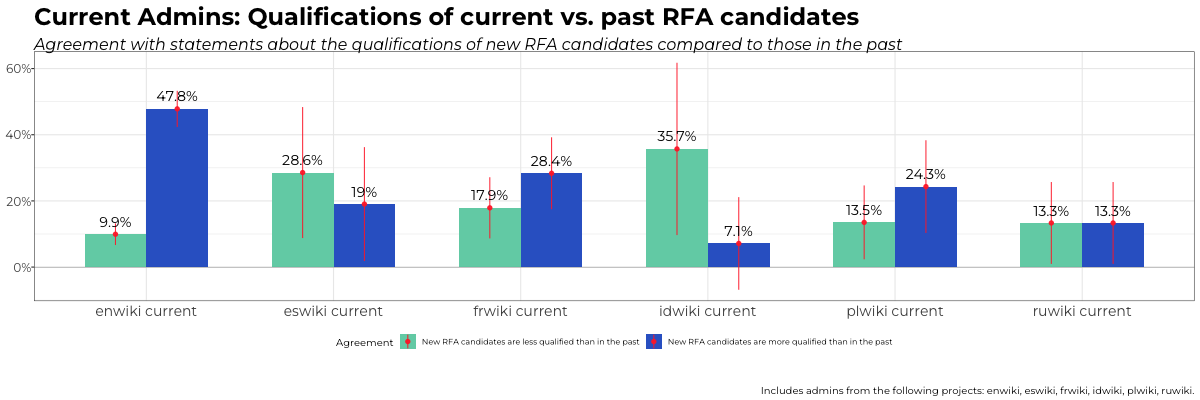}
  \caption{Current administrators' agreement with statements about the relative qualifications of current and former admin candidates on their project.}
  \Description{Barchart showing attitudes on qualifications of rfa candidates.}
  \label{fig:rfa_qualified}
\end{figure}

\begin{figure}[!ht]
  \includegraphics[width=\textwidth]{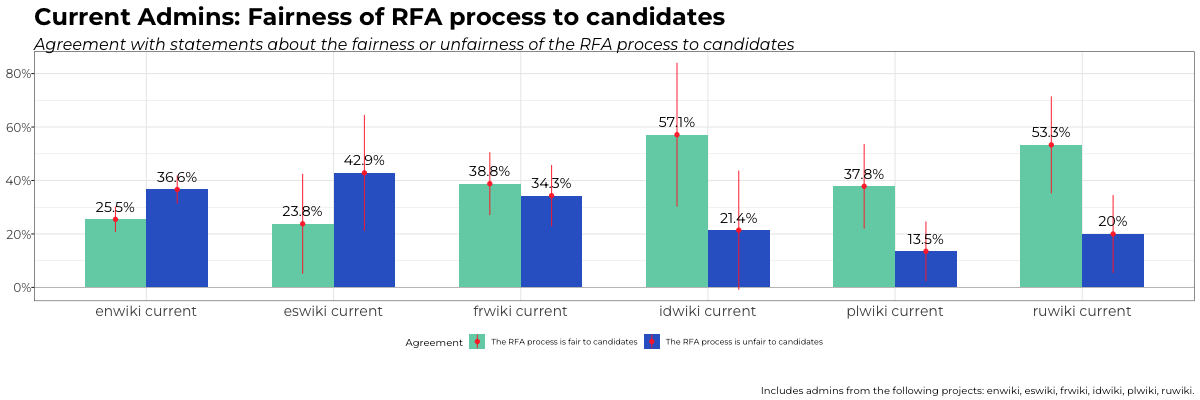}
  \caption{Current administrators' agreement with statements about the fairness of the RFA process on their project.}
  \Description{Barchart showing attitudes on the fairness of the rfa process.}
  \label{fig:rfa_fair}
\end{figure}

Nonetheless, current, former, and potential administrators consistently described the Request for Adminship (RFA) process as too demanding, highly stressful, and confusing. Over half of surveyed administrators (52.5\%) believe ``it is too difficult to become an administrator on Wikipedia,'' while only 7.6\% think the process is too easy. Results for individual language editions show the same overall pattern, with the exception of Indonesian Wikipedia (Appendix Figure~\ref{fig:rfa_ease})\footnote{Due to the small number of responses received for non-English language editions, drawing inferences about attitudes in these projects is frequently challenging, particularly because the sample sizes for individual survey questions are generally considerably smaller than those for the surveys as a whole due to survey attrition. These challenges are particularly acute for Indonesian Wikipedia (n=14). As a result, we generally present pooled estimates for current administrators (albeit with context about project-to-project variation). English Wikipedia administrators are necessarily influential on these estimates as they comprise 508 out of 805 (63\%) of total respondents. However, this slightly \textit{underrepresents} their share of the population of interest 838 out of 1285 (65\%) as shown in Table~\ref{tab:survey_combo_admins}.}. This is despite the fact that substantially more current administrators overall\footnote{By language edition, results are mixed, with English Wikipedia administrators viewing new RFA candidates as more qualified by a very large margin, but estimates that are statistically indistinguishable in the other projects (Appendix Figure~\ref{fig:rfa_qualified}.)} (40.6\%) believe ``new RFA candidates are more qualified than in the past'' than believe they are ``less qualified'' now (13.6\%). Current administrators overall are split on the fairness of the RFA process as 34.3\% consider it ``unfair to candidates'' while 29.3\% view it as fair\footnote{This division is reflected at the language edition level as well, with mixed results, although with substantial uncertainty outside of English Wikipedia (Appendix Figure~\ref{fig:rfa_fair})}. 
Descriptions of the process from former and potential administrators are almost universally negative: a former English Wikipedia admin described her RFA as ``the worst week of my life'', while a former Spanish Wikipedia administrator called it a ``mental battle''. 

\begin{figure}
  \includegraphics[width=\textwidth]{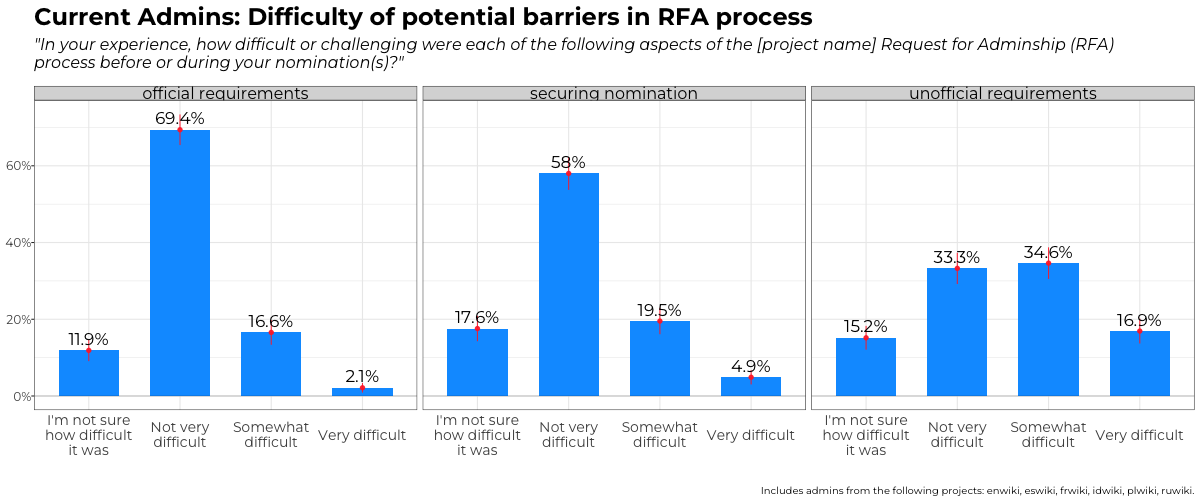}
  \caption{How difficult current Wikipedia administrators found three different aspects of the Request for Adminship (RFA) process on their projects. From a survey of current English, Spanish, French, Indonesian, Polish, and Russian Wikipedia administrators. Sample proportion estimates are shown with 95\%\ confidence intervals.}
  \Description{Barchart showing perceived difficulty of different aspects of the RFA process as described by Wikipedia admins}
  \label{fig:rfa-difficulty}
\end{figure}

\begin{figure}[!ht]
  \includegraphics[width=\textwidth]{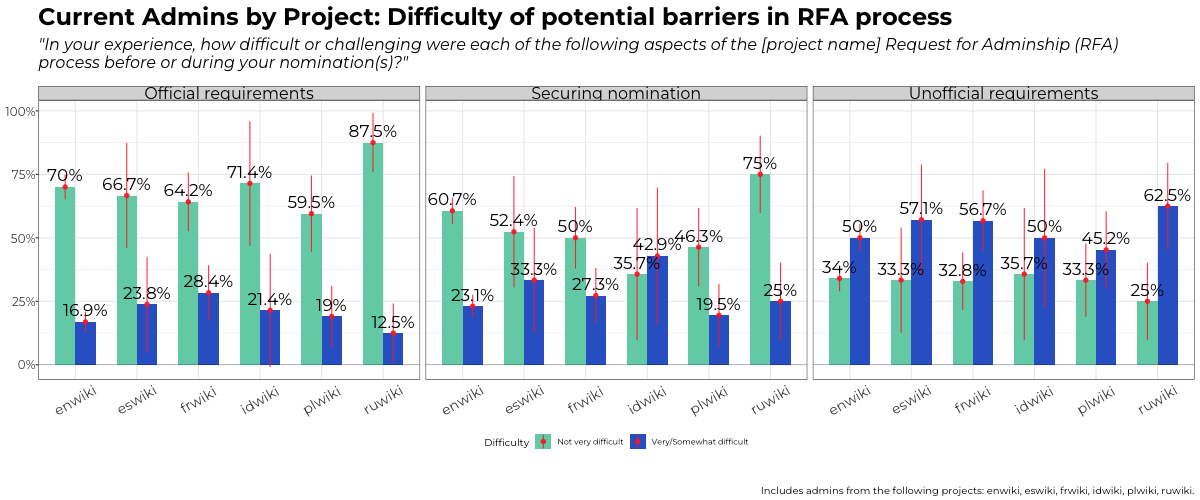}
  \caption{Current administrators' views on the relative difficulty of potential barriers in the RFA process.}
  \Description{Barchart showing attitudes on the difficulty of different aspects of the rfa process.}
  \label{fig:rfa_difficult_sample}
\end{figure}

In every project, unwritten norms about a candidate's reputation and standing in the community are crucial to RFA success. For example, as shown in Figure~\ref{fig:rfa-difficulty}, current Wikipedia administrators -- definitionally successful candidates -- regard the unofficial RFA requirements as the biggest barrier to a successful candidacy. Over half of current Wikipedia administrators surveyed say it was ``very'' (16.9\%) or ``somewhat'' difficult (34.6\%) to meet unofficial requirements. In contrast, a majority characterized meeting official requirements (69.4\%) or securing a nomination (58.0\%) as ``not very difficult''. These results are generally replicated in each language edition with admins much more likely to find meeting the unofficial requirements at least ``somewhat difficult'' compared to meeting official requirements or securing a nomination--although it is harder to draw conclusions for non-English Wikipedias (Appendix Figure~\ref{fig:rfa_difficult_sample}). Interviewees highlighted inconsistency in how requirements are interpreted during RFA votes, with one former English Wikipedia administrator outright stating there is ``no consistent standard'' for how RFAs are judged.

\subsubsection{Current administrators: prioritizing recruitment amidst pervasive conflict.}

\begin{figure}[!ht]
  \includegraphics[width=\textwidth]{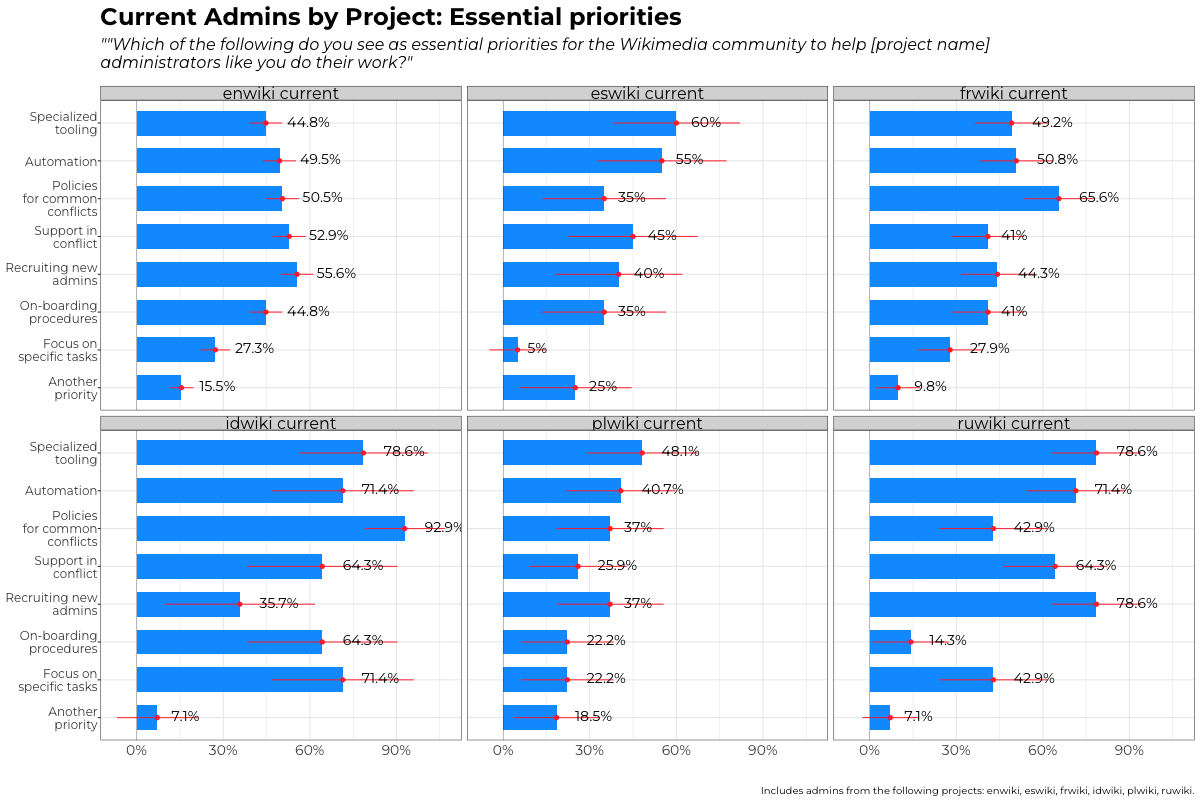}
  \caption{Proportion of current administrators naming a proposed improvement as an ``essential priority''.}
  \Description{Barchart showing admins' essential priorities.}
  \label{fig:current_priorities}
\end{figure}

Survey data highlight that recruitment of new administrators is one of the most essential priority areas for current admins: 52.2\% of respondents identified ``recruiting new admins'' as an essential area of focus, and a matching 52.2\% pointed to ``policies for common conflicts''. Nearly half (49.9\%) also cited the need for greater ``support in conflict''. At the level of specific language editions, these priorities are generally among the most commonly-cited by administrators in each project (Appendix Figure~\ref{fig:current_priorities}).  

\begin{figure}[!ht]
  \includegraphics[width=\textwidth]{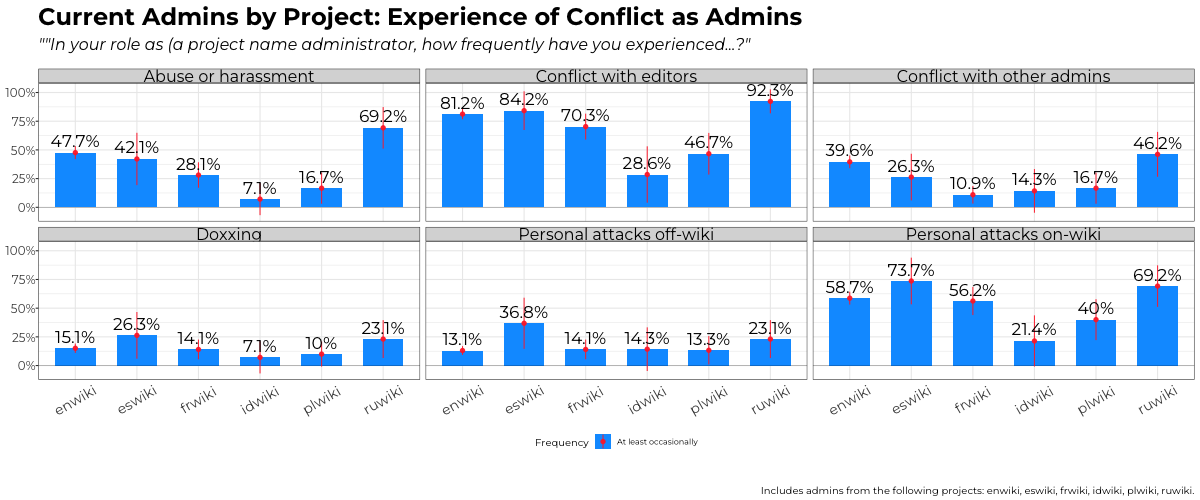}
  \caption{Proportion of current administrators experiencing different types of conflict ``occasionally'' or more.}
  \Description{Barchart showing admins' experiences of conflict.}
  \label{fig:conflict_combo}
\end{figure}

Conflict remains a pervasive obstacle in admin work. More than three-quarters (76.5\%) report occasional or frequent conflict with editors (59.9\% occasionally; 16.6\% frequently), and many experience harassment or challenging disputes\footnote{We observe largely consistent patterns across language editions and for different forms of conflict--although administrators on Indonesian and Polish Wikipedia generally report lower levels of experience with conflict (Appendix Figure~\ref{fig:conflict_combo}).}. This recurrent conflict adds to admins' burdens and compounds recruitment and retention difficulties.

Qualitative findings complement these findings, bringing up themes of conflict avoidance, calmness under pressure, and strong interpersonal skills when handling disputes or during the RFA process. Current administrators recounted changing their behavior prior to their RFA to ``avoid ruffling any feathers'', or to exercise ``extreme caution to be civil at all times''. Others described compromise and constructive engagement with editors and other admins as essential to navigating the stresses inherent in administrator roles.

Insights from interviews complement these findings, with both current and former admins emphasizing the need for impartiality, calmness under pressure, and strong interpersonal skills—especially during the RFA process and when managing disputes. Many described compromise and constructive engagement with editors and other admins as essential to navigating the stresses inherent in administrator roles.

\subsubsection{Underlying strength: Administrators are highly motivated and engaged.}
To complement our findings on recruitment challenges, we surveyed current administrators from six Wikipedia language editions to systematically examine what motivates them. Our analysis reveals that administrators are primarily driven by a sense of impact—helping others, combating misinformation, and upholding the collaborative Wikipedia ethos. Personal pride and responsibility in maintaining the quality of the platform are also central themes, as reflected in both survey responses and interviews:
\begin{quote}
    \textit{``I'm really passionate about this work. The maintenance keeps the quality of the site high... I want it to be a credible source of information in the world.''} -- current Spanish Wikipedia administrator
\end{quote}
and,

\begin{quote}
    \textit{``[I most enjoyed] seeing things that weren’t getting done, done.''} -- former English Wikipedia administrator
\end{quote}

Quantitatively, most administrators agree that they are proud of their work (81\%, 40\% of which reported strongly agree)\footnote{We observe consistently high agreement across all surveyed language editions.} and value contributing to the community good. Less emphasis is placed on recognition or social factors, such as peer status or forming friendships through admin work. The majority see their admin role as a meaningful way to ensure accurate and trustworthy information for all users. (See Fig.~\ref{fig:motivations_ranked} in Appendix for more details.)

And lastly, it is important to call out that very few interviewees reported negative sentiments related to their time as an administrator, even among those who had no interest in pursuing adminship again.

\begin{quote}
\textit{``[I was motivated by] my love for Wikipedia. Like, I love knowledge. I love disseminating knowledge. And I saw administrative work as part of that, to keep the information trustworthy and safe.''} -- former Indonesian Wikipedia administrator
\end{quote}


In summary, while barriers to recruitment remain formidable, our evidence suggests that current administrators are fueled by intrinsic motivation and a commitment to Wikipedia’s mission—a strength on which future recruitment and retention strategies can build.

\section{Discussion}
\label{sec:discussion}
This section includes a discussion of the results in the previous section, including an enumeration of the design implications stemming from them. We also discuss limitations, future directions for work in this area, and ethical considerations.

\subsection{Wikipedia is not experiencing a mass exodus of administrators, but concentrated decline on large projects poses systemic risks}
As a result of the current work, 
we are able to rigorously and systematically address past concerns that have been raised about declining numbers of admins on Wikipedia (e.g., \cite{signpost_recent_2015} and \cite{signpost_thirteen_2023}). As noted in the results, most Wikipedias have numbers of admins that are stable or growing; however, most larger language editions show declining numbers. This is important as it shows we are not observing a mass exodus. At the same time, if current trends on the projects with declining numbers persist, this may pose systemic risks to these projects because of the fewer numbers of users who are able to meet the moderation demands of the projects. Moreover, it could also mean that greater demands are put on fewer and fewer admins, which increases the risk of burnout. 

The current patterns of active admin presence on the projects has at least two additional implications for broader global knowledge production and governance. First, the largest language editions carry a disproportionate burden for global knowledge production given their size of content and editors as well as Wikipedia's important role in the knowledge ecosystem (see \textit{e.g.} \cite{piccardi2021value, penedo2024fineweb, gao2020pile, aly2021feverous}). As such their admin shortages can have outsized effects on content integrity and governance outside of Wikipedia. Secondly, results also demonstrate how stability (or even growth) of admin numbers in smaller language editions does not offset decline in larger projects\footnote{In part because each language edition is a separate project with its own set of policies and administrator candidacy guidelines.}. Thus, platform health must be judged by systemic risk, not an approach that considers averages. This has implications for prior work (e.g., \cite{butler_dont_2008, forte_decentralization_2009} ) that shows how growing communities struggle to handle increased coordination burdens via decentralization of power. Namely, in the case of Wikipedia, we show that decline is concentrated, not universal, and thus systemic risk depends on where decline occurs; on larger projects in this case.

\subsection{Recruitment shortfalls, not attrition, primarily drive decline and have implications for effective interventions}
For the language editions with declining admin populations, our results show that the decline currently results primarily from insufficient inflow of new admins, not an unusually high outflow, or rate of attrition. This has clear implications for interventions, and requires a diversification in attention from primarily focusing on retention strategies toward prioritizing building recruitment pipelines. This also contributes evidence that governance sustainability may in some cases be threatened by lack of inflow at least as much if not more than by outflow. This encourages a reframing of prior work on volunteer moderation, which has emphasized attrition and burnout as primary risks.~\cite{signpost_recent_2015,konieczny_volunteer_2018,schopke-gonzalez_why_2024} 

While existing moderation research often centers around burnout and attrition, our findings suggest more importance should be given to recruitment bottlenecks and barriers. The case for shifting attention to recruitment pipelines is made even stronger by our findings about the reasons for attrition. The factors that drive attrition are largely beyond the ability of individual communities to influence. Consider as one very clear example of this the common case of admins who have to leave their role due to changing life circumstances.


 Additionally, as we think about reducing social and technical barriers to entry, it is important to recognize the inertia towards maintaining the status quo\cite{steinsson_rule_2024}, even when the status quo is unsustainable. For example, despite attempts in 2015 and 2021, major reforms to English Wikipedia's RFA process finally took place from February 2024 to April 2025 \footnote{An overview of this reform is available at \url{https://en.wikipedia.org/wiki/Wikipedia:Requests_for_adminship/2024_review}}. The lengthy nature of the reform and the relatively minor changes made to the process prior to 2024, signals the difficulty that communities may face when trying to reform the candidacy process. An open question at the time of this work is what effects the 2024 RFA reform on English Wikipedia may have on admin numbers and the experiences of those candidates passing through the RFA process.


\subsection{Clarifying Adminship Pathways: Addressing Perceptual and Procedural Barriers}
Results of this investigation have shown that recruitment pipelines fail less because the current pool of potential administrators lack skills or eligibility for adminship, and more because pathways into adminship may be opaque, unattractive, or poorly supported. Recall that the current pool of potential admins often lack awareness of the RFA process, perceive candidacy as stressful (confirmed by current admins perceptions as well), and/or fear that admin work will crowd out their preferred forms of engagement (i.e., namely content editing in the context of Wikipedia). These deterrents are compounded by the limited opportunities for would-be admins to experience administrative tasks before candidacy, which widens the gap between outside perceptions and the realities of administrator day-to-day work.

This in turn can impact the sustainability of platform governance as communities may risk losing strong candidates not to formal skill gaps, but to cultural, procedural (e.g., formal or unstated candidacy guidelines), and perceptual barriers. Previous work (e.g., \cite{forte_decentralization_2009, butler_dont_2008}) has demonstrated how increasing policy complexity can discourage newcomer participation, and our findings extend these insights: namely a specific example of how critical failure points in governance pipelines can also include the candidacy process itself and the perception of admin work and awareness of the candidacy pipeline. Addressing these challenges requires clarifying candidacy pathways, reducing procedural barriers and unstated or informal requirements, and creating routes for learning about administrative work. 

\subsection{Design implications}
Given the challenges and opportunities highlighted in the previous sections, we now move to design implications that stem from this work. Addressing the challenges of administrator recruitment and retention is a socio-technical problem. The design implications we discuss below therefore span interventions at the level of workload, participation pathways, and cross-community learning.

\subsubsection*{Increasing the visibility of and participation pathways in administrator work} In this study we found that the awareness of adminship is limited among the current pool of potential administrators. We also found that the current pool of potential admins often lack awareness of the RFA process, perceive candidacy as stressful (confirmed by current admins perceptions as well), or fear that admin work will crowd out their preferred forms of engagement (i.e., namely content editing in the context of Wikipedia). We further know that learners need particular types of access for learning, with access to participation being top of the list. \cite{lave1991situated} We recommend Wikipedia languages to consider experimenting with increasing the visibility of admin work in their projects and providing more explicit pathways to do admin work for potential administrators who have not passed the RFA process (for different possible reasons) and can be interested in exploring some aspects of admin work. An example of this can be found on Portuguese Wikipedia, where \textit{rollbacker} group members -- a group with a much lower barrier to entry than administrators -- are allowed to block users with unconfirmed accounts, who have engaged in vandalism, for up to 24 hours.~\cite{ptwiki_reversores} The potential administrators can then progressively be exposed to other types of tasks and learning and through this ``ladder'' of contribution build confidence and skills required for being a successful admin. 

\subsubsection*{Investing in mentorship} Our study confirms prior research findings that the RFA process is significantly stressful or opaque for potential administrators in some Wikipedia languages. We also know from prior research in open source environments \cite{steinmacher2018let} as well as Wikipedia \cite{morgan2018evaluating} that newcomers can benefit from mentors and mentorship environments and dashboards. We encourage experimentation with mentorship for potential administrators to become administrators. We hypothesize that similar to other spaces, mentors for potential administrators can reduce the significant stress felt by potential administrators to participate in the RFA process.

\subsubsection*{Designing for cross-community learning and flexibility}
Finally, given the variation in processes, patterns and experimentation we observed across Wikipedia languages and considering that the focus of the administrators of each Wikipedia has traditionally been on the particular Wikipedia's needs and opportunities, we expect there to be promising opportunities for cross-community exchanges and knowledge sharing. For example, different communities of admins on different language editions have done their own experimenting around programming and policy to help facilitate the growth of administrator bodies.\footnote{For example, Indonesian Wikipedia has used a geographically-based chapter to encourage and mentor promising admin candidates, Polish Wikipedia has facilitated various administrator and other user group sharing sessions, and Ukrainian Wikipedia has introduced a temporary admin role and other admin-responsibility-dispersal approaches.}  At the same time, these sharing and learning opportunities must promote context-sensitive design by encouraging the building of policies and tools that are flexible enough to adapt to diverse local needs.   



\subsection{Limitations}

\subsubsection*{Case selection} 
This study focuses on administrators of a particular subsample of Wikipedia language editions -- those with relatively large administrator populations and relatively institutionalized practices, a group that largely, if not perfectly, coincides with the largest Wikipedia language editions by editors, pageviews, or content. This introduces a skew toward languages of European origin or with larger speaker bases. Given that incentives to contribute to Wikipedia have been found to depend causally upon community size \cite{zhang2011group}, findings from this study may be difficult to generalize to smaller language editions. Moreover, while we attempted to balance language editions selected for survey and interview data collection by MAA and editor numbers, web traffic, administrator policies and activity levels, and administrator candidacy requirements, we cannot rule out that these criteria may not include some other observable or unobservable dimension of variation that might be systematically related to the views and experiences of current, former, and potential administrators.

\subsubsection*{Survey and interview selection}
Similarly, non-response bias is a pervasive issue in survey research. Participation in research may be associated with attitudes, experiences, and demographic characteristics making survey-based estimates systematically biased even when, as in this study, researchers contact an entire population or randomly sample from it.\footnote{Note moreover that high response rates (such as those shown in Table~\ref{tab:survey_combo_admins}) do not necessarily indicate the absence of non-response bias nor do low response rates indicate its presence \cite{groves2006nonresponse,davern2013nonresponse,hendra2019rethinking}.} The challenge of non-response bias may be particularly acute in the case of Russian Wikipedia administrators. News reports of Russian government attacks on Russian Wikipedia suggest that administrators located within the Russian Federation may be much less likely to respond to survey research, an impression reinforced in interviews with the Wikimedia Foundation Human Rights Team. Researchers frequently attempt to address non-response bias through post-stratification weighting.\footnote{Weighting adjustments depend upon the \textit{strong and frequently violated} assumption that sample selection is ignorable conditional on observables.} However, because we lack auxiliary data on the ``true'' population parameters of Wikipedia administrators, the survey data in this study are unweighted. 

Interview data, particularly with former administrators, is subject to similar sample selection challenges. Because we only recruited former administrators who were 1) active Wikipedia editors or 2) remained involved with Wikimedia movement events, we were not \textit{e.g.}, able to interview former administrators who had completely cut ties with Wikipedia as a whole. Second, we excluded Wikipedia administrators who departed the role because they had been banned from their projects or by the Wikimedia Foundation as this would require designing a new interview protocol around a potentially hostile interviewee, and significantly increase risks to researchers. It appears likely that the extent and circumstances of former administrators' separation from Wikipedia would vary systematically with their experiences as an administrator.

\subsubsection*{Statistical power} 
Due to the small population of current Wikipedia administrators, statistical power is an unavoidable challenge for survey research on this population. This means that for current administrators, survey estimates at the language edition level outside of English Wikipedia typically come with substantial uncertainty. This is despite the fact that, as reported in Table~\ref{tab:survey_combo_admins}, >60\% of current administrators responded to the survey. For this reason, when reporting survey results for current administrators, we focus on pooled estimates, while providing additional context in the Appendix. 

\subsubsection*{Attrition forecast}
We found that the majority of the surveyed current administrators expect to be admins in two years and they rarely or never consider quitting. While the central finding in this research -- that declines are primarily driven by recruitment shortfalls and not attrition for large Wikipedias -- does not rest on a population-level attrition forecast, it is important to note some limitations with regards to these attrition findings given the importance of attrition and that we hope future research continues to focus on this topic as well. We recognize that survivorship and social desirability can push current admins' answers toward reporting rarely or never having considered leaving the admin role in two years. In addition, nonresponse can correlate with lower satisfaction or higher burnout which can in return introduce bias that can inflate the results related to intention to stay or deflate quit salience in a survivor sample.

\subsection{Future directions}
This study 
has presented a large-scale, multi-lingual, and multi-method examination of Wikipedia administrators, spanning 284 Wikipedia language editions. It substantially expands the limited body of scholarship on Wikipedia administrators (see Section~\ref{sec:related-work}). Given the diversity of organizational models across Wikipedia languages and the novel analytical dimensions introduced in this work, we recognize that our findings can inspire further avenues for research beyond those discussed below.

\subsubsection*{Focus on lower activity Wikipedia languages with adminship decline.} In this study, we chose to focus on Wikipedias with declining adminship trends that are large and highly active. We hypothesize that some of the challenges and opportunities for smaller Wikipedia languages with adminship decline to be different than the larger Wikipedias studied. Such studies can inform the research community about the generalizability of the findings in the current study.

\subsubsection*{Estimating optimal administrator capacity} A question we encountered in conducting this research is what constitutes an optimal number of administrators for a given Wikipedia language. Determining adequate administrative capacity is a complex, context-dependent problem potentially involving both endogenous and exogenous factors. Factors such as editing volume, content growth, vandalism patterns, community size, governance norms, availability of bots to assist with admin operations, the distribution of workload across current administrator bodies, the geopolitical environment a Wikipedia language is most associated with and more can affect this optimal number. While determining empirically validated models of optimal administrator actions or numbers is beyond the scope of the present study, we encourage future research in this space.

\subsubsection*{Studying interventions.} We expect that as Wikipedia language communities and particularly users with extended rights become more aware of the challenges and opportunities for growing the Wikipedia administrator pool, the communities apply interventions to change the decline patterns. We see effects of this in English Wikipedia\footnote{\url{https://en.wikipedia.org/wiki/Wikipedia:Requests_for_adminship/2024_review}}, for example. As communities design interventions, numerous research opportunities open. For example, research can study the effect of interventions through natural experiments, they can join forces with the local communities to design controlled experiments, or they can study the evolution of communities as they learn from these experiments through the choices they will make in policy and guideline updates. 

\subsubsection*{Power dynamics and recruitment.} When survey current Wikipedia administrators and interviewing current and former administrators, we found participants communicating a significant sense of responsibility towards Wikipedia and other editors. At the same time, we know from past research \cite{danescu_echoes_2012} that there are significant power dynamics around the RFA process. We encourage future research on how to bring these two realities together and design for recruiting systems that navigate the need for more transparent and less stressful pathways to recruitment while recognizing that there are some power dynamics at play. 

\subsubsection*{Open knowledge cross-platform comparisons.} We encourage future research to understand how different governance models in open knowledge systems mitigate or exacerbate recruitment bottlenecks and workload imbalances discovered as part of this research. Comparative studies across platforms such as OpenStreetMap, iNaturalist, Reddit, or other collaborative knowledge projects could illuminate how diverse organizational structures, decision-making processes, and community support mechanisms influence administrator sustainability and participation. Such cross-platform analyses will help identify transferable strategies and context-specific challenges in volunteer-driven governance.

\subsection{Ethical considerations}
Our work is subject to internal organizational review, and all our work was governed by relevant policies as well as guardrails of data publication guidelines and data retention guidelines. Survey data collection and retention was governed by specific privacy statement.\footnote{\url{https://foundation.wikimedia.org/wiki/Legal:Administrator_Experiences_2024_Survey_Privacy_Statement}} We only share project code containing queries and calculations of admins in the aggregate.

Part of our selection criteria for shortlisted Wikipedias was based on the privacy challenges of surveying or interviewing Wikipedia language versions with very small numbers of active administrators. Surveying or interviewing small groups of administrators drastically raises the possibility that individual participants may be identified on the basis of their responses. Therefore, we decided to exclude these Wikipedia editions from qualitative data gathering. All interviews and surveys were accompanied by privacy statements. Interview participants were anonymized and assigned a participant number. As described in the privacy statements, to protect research participants, we purged raw data after 90 days.

We will not share code that we have used to gather the administrator log data although we have explained sufficient information in the Methodology section for reproducibility. While the administrator log data is already public, making logs associated with administrator accounts at scale more easily available can risk Wikipedia administrators.


\section{Conclusion}
\label{sec:conclusion}
Wikipedia administrators are indispensable to the platform's ability to provide free, verifiable knowledge globally, yet their dynamics and challenges have been largely understudied. This work addresses that critical gap through the first systematic, multi-language analysis of administrator trends, motivations, and experiences.

Our findings reveal a critical two-sided trend: while adminship is stable or growing across the majority of 284 Wikipedia languages studied, a significant decline persists in most of the 21 highly active Wikipedias since 2018. We conclusively demonstrate that this decline is driven primarily by insufficient recruitment, not by increased attrition. Our mixed-methods investigation, integrating administrator logs, over 3000 surveys, and 12 interviews, pinpointed key barriers for potential administrators, including limited awareness of the role, ambiguous candidacy processes, and concerns about the nature of administrative work.

Crucially, our study also highlights a significant underlying strength: current administrators remain highly motivated and deeply committed to Wikipedia's mission, driven by a profound sense of impact and responsibility.

Ultimately, these findings underscore that the sustainability of Wikipedia's volunteer moderation hinges on strengthening its administrator recruitment pipeline. We conclude with actionable recommendations for Wikipedia communities to clarify pathways, reduce procedural obstacles, and foster a culture that supports new administrators, design recommendations, as well as  suggestions for future research to further advance this understanding.

\subsection*{Acknowledgments}
This work was not possible without collaboration with multiple individuals and support from other Wikimedia Foundation teams. A special thanks to Bethany Gerdemann for assisting with project logistics, including recruitment and language support, as well as helping with the initial write-up of results. Thanks also to Daisy Chen for helping us initially process and summarize results. In addition, we would like to acknowledge support we received from the Wikimedia Foundation Research Team, for feedback, the Movement Communications Team, for assistance with communications and outreach, and the Trust and Safety and Human Rights teams for feedback to help us ensure protection of research participants and Wikipedia administrators. A special thank you to all our research participants, editors and administrators who contributed comments, feedback, and suggestions. Finally, thank you to all Wikipedia administrators for their continued efforts to make Wikipedia a trusted and reliable source for readers worldwide.

We also disclose the use of large language model tools to prepare this paper. More specifically, we have used Gemini and ChatGPT for plot formatting suggestions, suggestions for cleaning survey data, and code (SQL, Ptyhon, and R codes specifically). We also used Gemini and ChatGPT to improve the flow of this paper including receiving recommendations on combining sections, or re-writing certain sections for clarity. We have used these technologies as assistive, have reviewed the output of them carefully as relevant, and accept full responsibility for what we present in this paper.


\bibliographystyle{ACM-Reference-Format}
\bibliography{bibliography} 

\section{Appendix }
This appendix contains tables and figures that provide additional supporting evidence for claims made in the paper.

 \begin{figure}[!ht]
  \includegraphics[width=\textwidth]{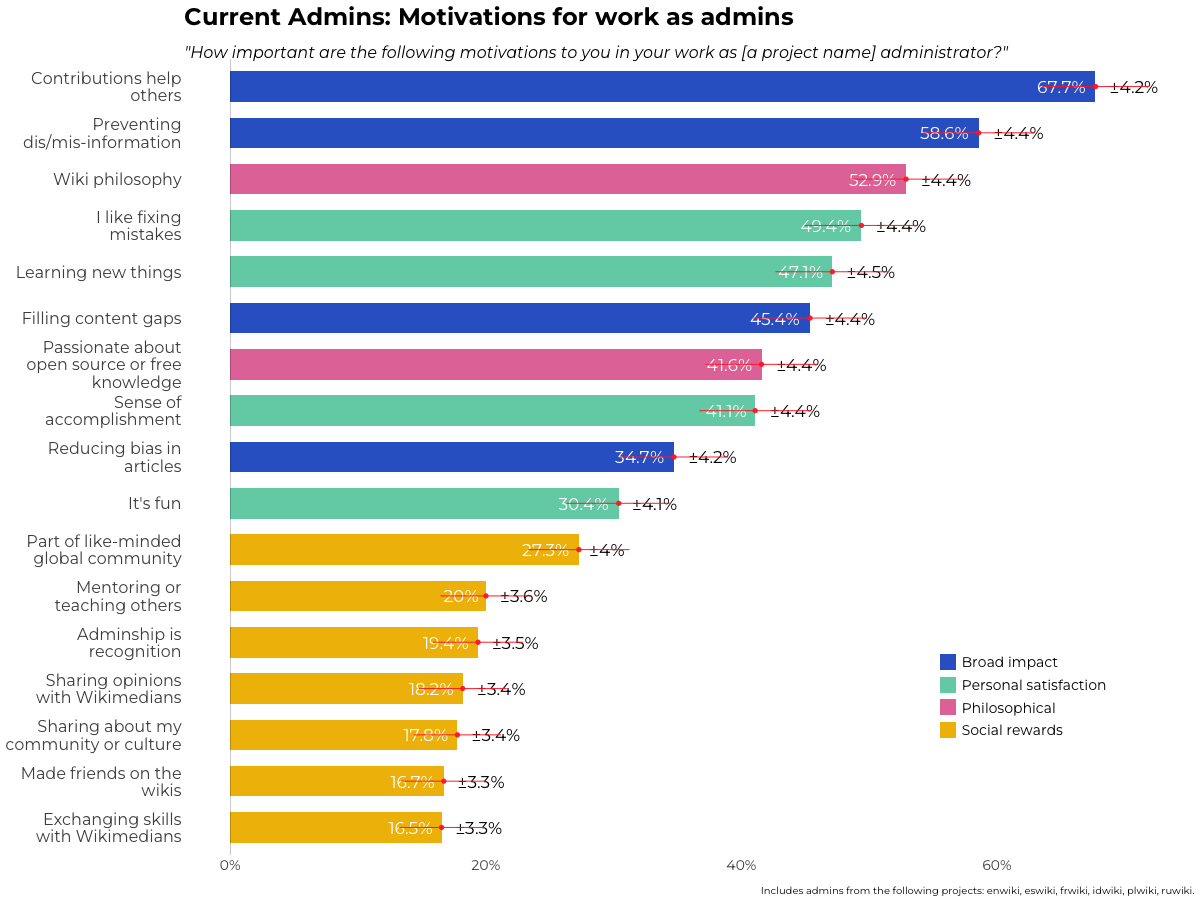}
  \caption{Proportion of current administrators who rate motivations as ``very important'' to their work as Wikipedia administrators. Motivations are categorized as relating to ``broad impact'' on Wikipedia, ``personal satisfaction'' in their work, ``philosophical'' commitment, and ``social rewards''. From a survey of current English, Spanish, French, Indonesian, Polish, and Russian Wikipedia administrators. Sample proportion estimates are shown with 95\%\ confidence intervals.}
  \Description{Barchart showing the share of motivations rated as ``very important'' by Wikipedia administrators.}
  \label{fig:motivations_ranked}
\end{figure}


\begin{figure}
    \centering
    \includegraphics[width=1\linewidth]{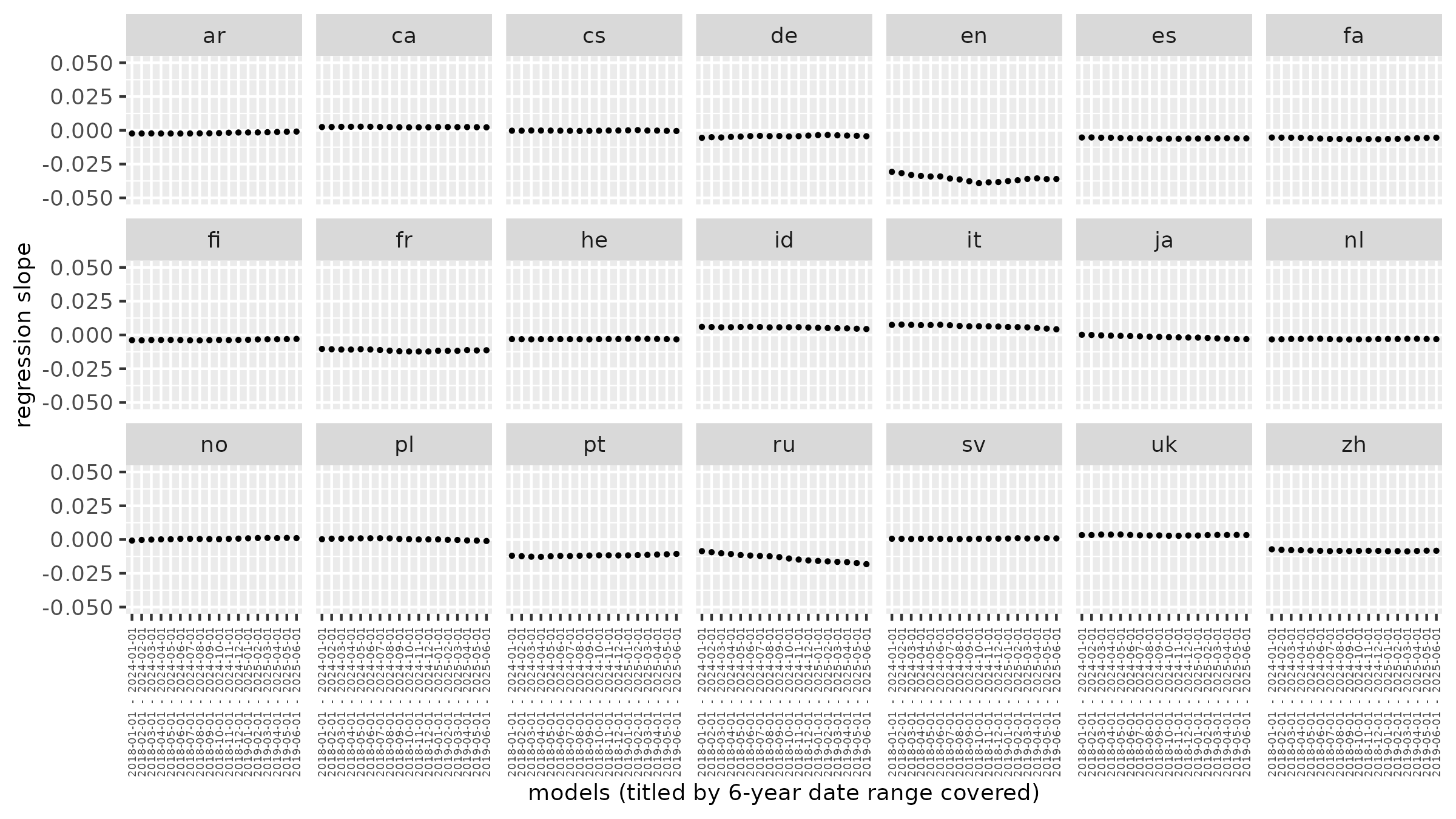}
    \caption{Regression slopes using repeating measures with a sliding scale. The x-axis shows the model name (titled by the date range covered) starting with the model covering Jan 2018-Jan 2024, and ending with the model covering June 2019-June 2025; the y-axis is regression slope value.}
    \Description{Regression slopes using repeating measures with a sliding scale. The x-axis shows the model name (titled by the date range covered) starting with the model covering Jan 2018-Jan 2024, and ending with the model covering June 2019-June 2025; the y-axis is regression slope value.}
    \label{fig:regression-slope-comparisons}
\end{figure}

\begin{table}
\centering
\caption{List of 21 shortlisted Wikipedia languages with large number of administrator, and the following statistics for the years 2018 and 2024: overall size rank (\cite{movement_insights_wiki_2024}), average monthly editors (\cite{movement_insights_wiki_2024}), average monthly edits (\cite{wikistats_general}), and average monthly admin actions (\cite{logging_table}). Overall size rank is a composite ranking of wikis by monthly active editors and monthly unique devices (produced by taking the geometric mean of those two values).}
\label{tab:wiki-activity-2018-2024}
\resizebox{\linewidth}{!}{%
\begin{tblr}{
  column{2} = {c},
  column{6} = {c},
  cell{1}{2} = {c=4}{},
  cell{1}{6} = {c=4}{},
  vline{2-3} = {1}{},
  vline{2,6} = {2-23}{},
  hline{1,24} = {-}{0.08em},
  hline{3} = {-}{0.05em},
}
                 & 2018 stats            &                            &                          &                                    & 2024 stats             &                            &                          &                                     \\
Wikipedia        & {Overall \\size rank} & {Avg. \\monthly \\editors} & {Avg. \\monthly \\edits} & {Avg. \\monthly \\admin \\actions} & {Overall \\size\\rank} & {Avg. \\monthly \\editors} & {Avg. \\monthly \\edits} & {Avg. \\monthly \\admin. \\actions} \\
English          & 1                     & 135,389                    & 4,561,276                & 103,411                            & 1                      & 123,996                    & 5,987,597                & 209,775                             \\
Spanish          & 2                     & 18,174                     & 662,100                  & 10,926                             & 2                      & 14,976                     & 657,603                  & 149,830                             \\
German           & 3                     & 20,389                     & 876,412                  & 75,543                             & 3                      & 17,891                     & 817,833                  & 12,773                              \\
Japanese         & 4                     & 13,474                     & 348,609                  & 4,526                              & 4                      & 13,147                     & 365,594                  & 4,980                               \\
French           & 5                     & 19,329                     & 905,468                  & 16,267                             & 5                      & 17,606                     & 863,283                  & 14,518                              \\
Russian          & 6                     & 11,718                     & 561,290                  & 282,154                            & 6                      & 10,021                     & 569,686                  & 231,842                             \\
Italian          & 8                     & 8,840                      & 679,955                  & 12,501                             & 10                     & 7,881                      & 460,028                  & 11,717                              \\
Chinese          & 9                     & 8,272                      & 399,237                  & 6,016                              & 8                      & 7,353                      & 439,135                  & 14,573                              \\
Portuguese       & 10                    & 6,608                      & 234,359                  & 9,414                              & 11                     & 8,931                      & 161,602                  & 4,673                               \\
Polish           & 11                    & 4,383                      & 322,011                  & 16,221                             & 14                     & 4,576                      & 261,348                  & 7,443                               \\
Arabic           & 12                    & 5,096                      & 472,897                  & 4,626                              & 13                     & 4,322                      & 269,481                  & 5,506                               \\
Dutch            & 13                    & 4,106                      & 178,193                  & 23,513                             & 16                     & 3,723                      & 154,684                  & 10,176                              \\
Persian          & 14                    & 4,892                      & 273,235                  & 8,751                              & 12                     & 5,536                      & 190,230                  & 5,350                               \\
Indonesian       & 15                    & 2,657                      & 78,946                   & 2,258                              & 15                     & 2,885                      & 132,651                  & 2,573                               \\
Ukrainian        & 19                    & 2,861                      & 187,606                  & 2,041                              & 19                     & 3,336                      & 241,794                  & 3,408                               \\
Swedish          & 20                    & 2,747                      & 130,503                  & 7,552                              & 24                     & 2,056                      & 184,217                  & 2,322                               \\
Czech            & 23                    & 2,123                      & 83,520                   & 1,905                              & 23                     & 2,355                      & 82,354                   & 1,341                               \\
Hebrew           & 24                    & 2,986                      & 192,025                  & 3,500                              & 22                     & 3,340                      & 208,097                  & 4,000                               \\
Finnish          & 26                    & 1,774                      & 64,698                   & 779                                & 28                     & 1,608                      & 64,616                   & 661                                 \\
Norwegian Bokmål & 27                    & 1,515                      & 79,639                   & 1,736                              & 34                     & 1,112                      & 59,731                   & 952                                 \\
Catalan          & 32                    & 1,476                      & 99,360                   & 983                                & 36                     & 1,210                      & 128,874                  & 848                                 
\end{tblr}
}
    \label{tab:wiki-activity-2018-2024}
    \Description{List of 21 shortlisted Wikipedia languages with large number of administrator, and the following statistics for the years 2018 and 2024}
\end{table}

\begin{table}
    \begin{tabular}{p{0.2\textwidth} | p{0.7\textwidth}}
        \hline
        Interview section & Sample questions\\
        \hline
        Introduction & ``How long have you been contributing to Wikipedia?'', ``What Wikipedia project do you consider your `home' or primary wiki?'', ``How did you learn about administrators on Wikipedia?''\\
        Gaining admin rights & ``Why did you seek out admin rights?'', ``Looking back, at the time that you applied to be an admin, do you think you were a suitable candidate? Why/why not?'', ``Did you have any help when you first applied for admin rights?'', ``What help or advice did you receive? Who provided it? What was most useful?''\\
        Perception of social status & ``How would you describe the differences between editors and administrators, in terms of responsibilities or focuses?'', ``Do you believe you were treated differently after becoming an administrator, and if so, how?'', ``Would you want to seek out additional user rights? What would they be, and why?''\\
        Attitudes towards RFA & ``Do you think it’s harder now to become an admin, compared to when you became an admin?'', ``Would you advise someone to apply for admin rights now?'' \\
        Experiences and motivations as admin & ``What kinds of admin work do you currently do?'', ``What, if any, day-to-day experiences make you feel motivated to continue admin work?'', ``What, if any, day-to-day experiences make you feel most demotivated?'' \\
        Conflicts & ``Have you ever gotten into a conflict with other editors? With other admins?'', ``How are conflicts usually resolved, if they are resolved?''\\
        Losing admin rights & ``What caused you to lose your admin rights?'', ``How would you explain your reasons for leaving admin work?'', ``How did this affect your general activity on the project?'', ``Is there anything that would make you reconsider being an admin on Wikimedia projects?''\\
        Closing & ``Looking back, what are your impressions on your time spent as an administrator?'', ``Overall, is your perception of administrators on Wikipedia positive or negative?''\\
        \hline
    \end{tabular}
    \caption{A sample of questions used in semi-structured interviews with current and former administrators, organized by theme.}
    \Description{A sample of questions used in semi-structured interviews with current and former administrators, organized by theme.}
    \label{tab:interview_questions}
\end{table}

\end{document}